\documentclass[twocolumn,bibnotes,superscriptaddress,showpacs,amsmath,amssymb,aps,prl]{revtex4-1}

\usepackage{graphicx}
\usepackage{amsmath}
\usepackage{amssymb}
\usepackage{color}
\usepackage{comment}
\usepackage[normalem]{ulem}
\usepackage{bm}
\usepackage[colorlinks=true,linkcolor=blue,citecolor=blue]{hyperref}

\begin{document}

\title{Anisotropic Enhancement of Lower Critical Field in Ultraclean Crystals of Spin-Triplet Superconductor UTe$_2$}

\author{K.~Ishihara}\email{k.ishihara@edu.k.u-tokyo.ac.jp}
\author{M.~Kobayashi}
\author{K.~Imamura}
\affiliation{Department of Advanced Materials Science, University of Tokyo, Kashiwa, Chiba 277-8561, Japan}

\author{M.~Konczykowski}
\affiliation{Laboratoire des Solides Irradi{\' e}s, CEA/DRF/IRAMIS, Ecole Polytechnique, CNRS, Institut Polytechnique de Paris, F-91128 Palaiseau, France}

\author{H.~Sakai}
\author{P.~Opletal}
\author{Y.~Tokiwa}
\author{Y.~Haga}
\affiliation{Advanced Science Research Center, Japan Atomic Energy Agency, Tokai, Ibaraki 319-1195, Japan}

\author{K.~Hashimoto}
\author{T.~Shibauchi}\email{shibauchi@k.u-tokyo.ac.jp}
\affiliation{Department of Advanced Materials Science, University of Tokyo, Kashiwa, Chiba 277-8561, Japan}

\begin{abstract}
The paramagnetic spin-triplet superconductor UTe$_2$ has attracted significant attention because of its exotic superconducting properties including an extremely high upper critical field and possible chiral superconducting states. Recently, ultraclean single crystals of UTe$_2$ have become available, and thus measurements on these crystals are crucial to elucidate the intrinsic superconducting properties. Here, we report the thermodynamic critical field $H_{\rm c}$, the lower critical field $H_{\rm c1}$, and the upper critical field $H_{\rm c2}$ at low fields of these high-quality single crystals. From the comparison of the anisotropies in $H_{\rm c1}$ and $H_{\rm c2}$, we find that the experimental $H_{\rm c1}$ values with the magnetic field along $b$- and $c$-axes are anomalously enhanced, showing unusual low-temperature upturns. We propose an effect of the strong Ising-like ferromagnetic fluctuations on the vortex line energy as the origin of the anisotropic enhancement of $H_{\rm c1}$.
\end{abstract}

\maketitle

The superconducting state of UTe$_2$ has been intensively investigated because of its spin-triplet nature with a paramagnetic normal state\,\cite{Ran2019S,Aoki2022R}. The spin-triplet pairing state induces an extremely high upper critical field ($H_{\rm c2}$) beyond the Pauli limit\,\cite{Ran2019S,Ran2019NP} and multiple superconducting phases under hydrostatic pressure or magnetic field\,\cite{Braithwaite2019,Thomas2020,Aoki2020,Kinjo_arxiv,Rosuel_arxiv,Sakai_arxiv}. Furthermore, chiral spin-triplet states are discussed in scanning tunneling microscopy\,\cite{Jiao2020}, polar Kerr effect\,\cite{Hayes2021}, surface impedance\,\cite{Bae2021}, and penetration depth studies\,\cite{Ishihara_arxiv}. As for the pairing interactions, the anisotropic $H_{\rm c2}$ and reentrant superconductivity\,\cite{Ran2019NP,Knebel2019} suggest that ferromagnetic fluctuations are at play analogous to the ferromagnetic superconductors\,\cite{Aoki2019}, which is supported by the nuclear magnetic resonance (NMR) and muon spin rotation ($\mu$SR) measurements\,\cite{Tokunaga2019,Sundar2019,Tokunaga2022}. On the other hand, recent neutron scattering studies revealed the presence of antiferromagnetic fluctuations related to superconductivity\,\cite{Duan2020,Knafo2021PRB,Duan2021}. Thus, the nature of magnetic fluctuations and their relationship to superconductivity are still unclear.

What has complicated systematic understandings of the superconducting state in UTe$_2$ are the inhomogeneity of crystals and the presence of magnetic impurities. Single crystals grown by the conventional chemical vapor transport (CVT) method sometimes show double superconducting transitions likely induced by inhomogeneity\,\cite{Hayes2021,Thomas2021,Rosa2022} and a small increase in magnetic susceptibility at low temperatures, which reflects the presence of U vacancies acting as magnetic impurities\,\cite{Ran2019S,Rosa2022,Knafo2021CP,Tokunaga2022,Haga2022}. Recently, Sakai {\it et al.} developed another crystal growth method called the molten salt flux (MSF) method and succeeded in obtaining high-quality single crystals of UTe$_2$ with a transition temperature $T_{\rm c}=2.1$\,K, a residual resistivity ratio as high as 1,000, and low magnetic impurity density\,\cite{Sakai2022}. In these ultraclean crystals, novel superconducting and normal state properties have been reported in addition to the observation of de Haas-van Alphen oscillations\,\cite{Sakai_arxiv,Tokiwa_arxiv,Aoki2022}. Therefore, measurements on the high-quality single crystals are crucial to clarify the superconducting nature in UTe$_2$.

In this study, we investigate the thermodynamic critical field ($H_{\rm c}$), the lower critical field ($H_{\rm c1}$), and $H_{\rm c2}$ in high-quality MSF single crystals. Regarding the Ginzburg-Landau (GL) theory, the anisotropies in $H_{\rm c1}$ and $H_{\rm c2}$ should be opposed because the critical fields satisfy $H_{\rm c1}H_{\rm c2}=H_{\rm c}^2 (\ln \kappa +0.5)$, where $\kappa$ is the GL parameter, and $H_{\rm c}$ is independent of field directions. In contrast, we find that the anisotropy of $H_{\rm c1}$ does not follow the expectations from that of $H_{\rm c2}$. Quantitatively, the above GL relation holds for ${\bm H} \parallel {\bm a}$, while it is obviously violated for ${\bm H} \parallel {\bm b}$. Furthermore, $H_{\rm c1}$ for ${\bm H} \parallel {\bm b}$ and ${\bm H} \parallel {\bm c}$ show unusual upturns at low temperatures. To explain these anomalous behaviors of the critical fields, we propose an anisotropic enhancement of $H_{\rm c1}$ induced by the Ising-like ferromagnetic fluctuations in UTe$_2$.

High-quality single crystals are grown by the MSF method as described in Ref.\,\cite{Sakai2022}. Crystals \#A1 and \#A2 in this study are the same samples as crystal \#A1 in Ref.\,\cite{Ishihara_arxiv} and crystal \#M7-1 in Ref.\,\cite{Sakai2022}, respectively, and crystal \#A3 for the $H_{\rm c2}$ measurement is picked up from the same batch with the crystal used in the de Haas-van Alphen experiments\,\cite{Aoki2022}. Crystals \#A1 and \#A2 have a cuboid shape with dimensions $455 \times 250 \times 95$\,$\mu$m$^3$ and $1400 \times 285 \times 165$\,$\mu$m$^3$, respectively.
$H_{\rm c}$ is calculated from the specific heat measured by the long-relaxation method where a Cernox resistor is used as a thermometer, a heater, and a sample stage\,\cite{Tanaka2022}.
$H_{\rm c1}$ is measured by a miniature Hall-sensor array probe tailored in a GaAs/AlGaAs heterostructure\,\cite{Shibauchi2007,Okazaki2009,Okazaki2010,Putzke2014}. The distance of neighboring sensers is 20\,$\mu$m so that we can measure the local magnetic induction and the first flux penetration field $H_{\rm p}$ near crystal edges.
$H_{\rm c2}$ is estimated from specific heat data $C(T)$ near $T_{\rm c}$ measured in the Physical Property Measurement System (Quantum Design). The same single crystal piece appropriately shaped relative to the principal axes was used for all the field directions relative to the crystal orientation.


\begin{figure}[t]
    \includegraphics[width=0.7\linewidth]{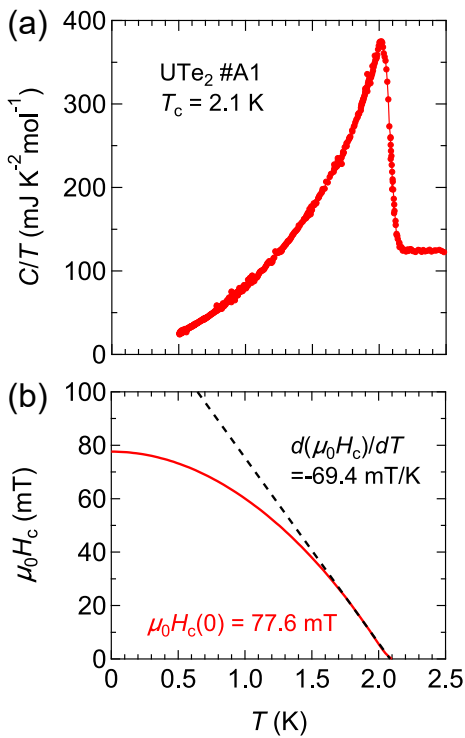}
    \caption{(a)\,Specific heat $C$ divided by temperature $T$ as a function of temperature measured in crystal \#A1. (b)\,Temperature dependence of the thermodynamic critical field calculated from the specific heat data. The black broken line represents the initial slope of $H_{\rm c}$ estimated near $T_{\rm c}$.}
    \label{Fig1}
\end{figure}

Figure\,\ref{Fig1}(a) shows the temperature dependence of $C/T$ of crystal \#A1. A sharp peak at 2.1\,K and a small residual value at low temperatures confirm that this crystal is ultraclean.  $H_{\rm c}$ can be calculated via $H_{\rm c}^2(T)=2\mu_0 \int_T^{T_{\rm c}} (S_{\rm n}-S_{\rm sc}) dT$, where $\mu_0$ is the permeability of vacuum and $S_{\rm sc}$ and $S_{\rm n}$ are the electronic entropy in the superconducting and normal states, respectively, calculated from the specific heat. Note that, to satisfy the entropy balance, we assume linear temperature dependence of the electronic contribution of $C/T$ in the normal state below $T_{\rm c}$ (see Supplemental Material). The obtained $\mu_0 H_{\rm c}(0) = 77.6$\,mT shown in Fig.\,\ref{Fig1}(b) is higher than the reported value in a low-$T_{\rm c}$ crystal\,\cite{Paulsen2021}, reflecting the higher quality of our crystal.

\begin{figure*}[t]
    \includegraphics[width=0.8\linewidth]{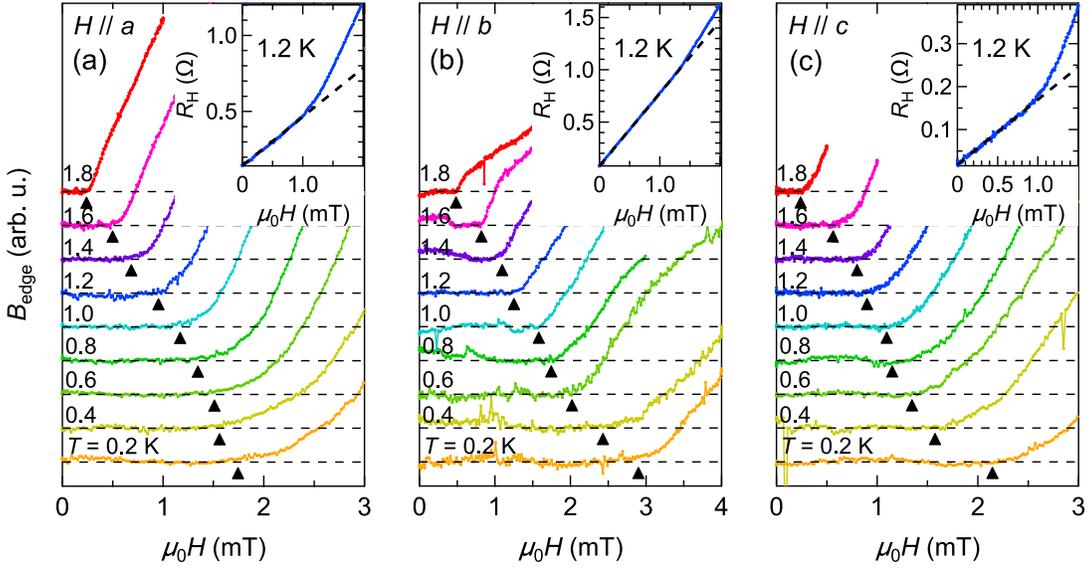}
    \caption{(a-c)\,The insets show a typical magnetic field dependence of the Hall resistivity near a crystal edge with the magnetic field along each crystallographic axis. The black broken line is the linear fitting of the data at the low-field region. The main panels show the local magnetic induction measured in crystal \#A1 with the magnetic field along $a$-, $b$-, and $c$-axes, respectively. The black rectangles are the first penetration field $H_{\rm p}$. The data are shifted vertically for clarity, and the black broken lines show $B_{\rm edge}=0$ for each measurement temperature.}
    \label{Fig2}
\end{figure*}

Figure\,\ref{Fig2} represents external magnetic-field dependence of the local magnetic induction $B_{\rm edge}$ near the crystal edges of crystal \#A1. The typical behaviors of the Hall resistance of the Hall-array probe near the crystal edges are shown in the insets in Fig.\,\ref{Fig2}. The linear dependence at the low field region originates from the finite distance between the Hall sensor and the crystal. By subtracting the linear function, we can derive the local magnetic induction $B_{\rm edge}$. At lower fields, $B_{\rm edge}=0$ because of the Meissner state. As the magnetic field increases, the $B_{\rm edge}$ value deviates from zero at $H_{\rm p}$ depicted as black triangles in Fig.\,\ref{Fig2}. We found that the $H_{\rm p}$ value does not depend on the crystal positions (see Supplemental Material), confirming that our measurements are free from surface pinning or geometrical barriers of superconducting vortices. Here, we note that $H_{\rm p}$ is lower than the bulk $H_{\rm c1}$ because of the demagnetization effect. The relation between $H_{\rm p}$ and $H_{\rm c1}$ has been precisely studied\,\cite{Brandt1999}, and we can evaluate $H_{\rm c1}$ from $H_{\rm p}$ through
\begin{equation}
  H_{\rm c1}=\frac{H_{\rm p}}{\tanh \sqrt{0.36t/w}},
  \label{Hc1_eq}
\end{equation}
where $t$ and $w$ are the sample thickness and width, respectively.

\begin{figure*}[t]
    \includegraphics[width=0.7\linewidth]{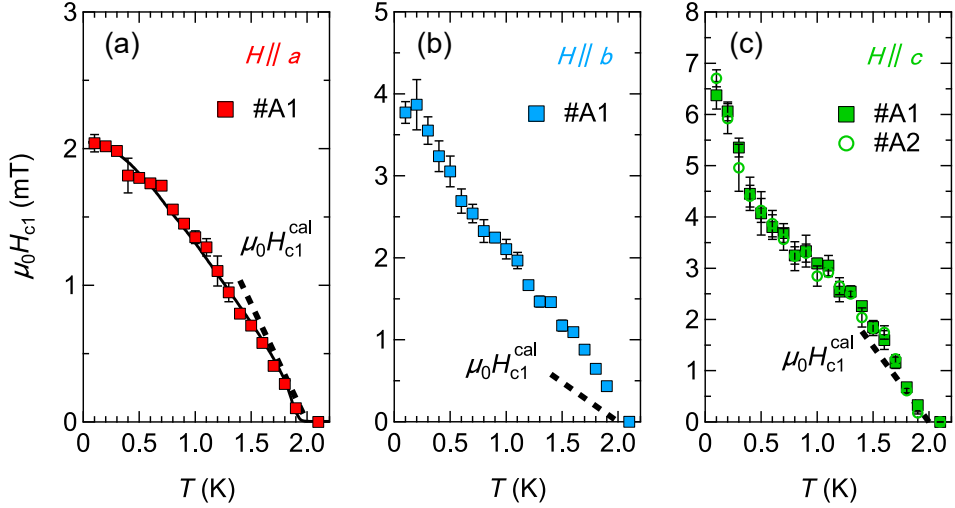}
    \caption{(a-c)\,Lower critical field as a function of temperature with the magnetic field along $a$-, $b$-, and $c$-axes, respectively, evaluated from Eq.\,(\ref{Hc1_eq}). The black broken lines represent the initial slope of the lower critical field expected from the usual GL relations. The black solid line in (a) is obtained by multiplying the normalized superfluid density $(\lambda_b^2(0)/\lambda_b^2(T)+\lambda_c^2(0)/\lambda_c^2(T))/2$ by a constant value calculated from the temperature dependence of the anisotropic penetration depth measurements\,\cite{Ishihara_arxiv}. The filled squares and open circles are the data obtained in crystals \#A1 and \#A2, respectively.}
    \label{Fig3}
\end{figure*}

The obtained $H_{\rm c1}^i$ with the magnetic field along $i$-axis is shown in Fig.\,\ref{Fig3}. Considering the usual relation
\begin{equation}
  \mu_0 H_{\rm c1} = \frac{\phi_0}{4\pi \lambda^2} (\ln \kappa+0.5),
\end{equation}
$H_{\rm c1}(T)$ is expected to scale with the normalized superfluid density $\rho_s(T) \propto \lambda^{-2}(T)$ when the temperature dependence of the GL parameter, $\kappa$, is small. The black solid line in Fig.\,\ref{Fig3}(a) is the normalized superfluid density for ${\bm H} \parallel {\bm a}$, $\rho_s^a = (\lambda_b^2(0)/\lambda_b^2(T)+\lambda_c^2(0)/\lambda_c^2(T))/2$, calculated from the anisotropic penetration depth measurements\,\cite{Ishihara_arxiv}. It is obvious that $H_{\rm c1}$ and $\rho_s^a$ show similar $T$ dependence, confirming the consistency of $H_{\rm c1}$ and penetration depth measurements. On the other hand, the $H_{\rm c1}$ data along $b$- and $c$-axes show an anomalous concave temperature dependence around 0.5\,K, which is not detected in the penetration depth measurements. The origin of this unusual $T$ dependence will be discussed later. We note that, although the absolute value of $H_{\rm c1}^c$ in crystal \#A1 may have an estimation error of the demagnetization effect because of its plate-like shape, we obtain very similar $H_{\rm c1}^c (T)$ in another crystal \#A2 [Fig.\,\ref{Fig3}(c)], confirming the correct estimation of $H_{\rm c1}^c$ values in these crystals.

\begin{figure}[t]
    \includegraphics[width=0.9\linewidth]{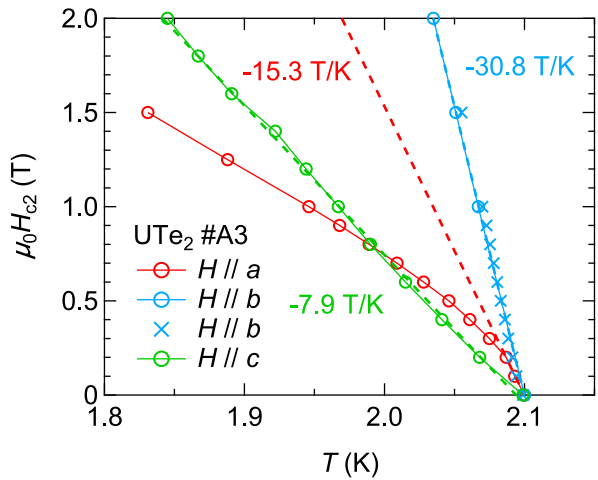}
    \caption{Upper critical field in the low-field region with the magnetic field along $a$-axis (red), $b$-axis (blue), and $c$-axis (green) in crystal \#A3. The data are obtained by specific heat measurements on a high-quality crystal with $T_{\rm c}=2.1$\,K. The broken lines are the initial slopes obtained by the linear fitting of the data near $T_{\rm c}$.}
    \label{Fig4}
\end{figure}

First, we compare the anisotropies in $H_{\rm c1}$ and $H_{\rm c2}$. As mentioned before, the GL theory implies that the anisotropies in these critical fields are opposite as long as we discuss the initial slope near $T_{\rm c}$ where the GL formalism is valid. Figure\,\ref{Fig4} depicts $H_{\rm c2}^i(T)$ with the magnetic field along $i$-axis near $T_{\rm c}=2.1$\,K. The slope of $H_{\rm c2}(T)$ along $a$-axis significantly changes with decreasing temperature in the low-field region, which has been already reported in previous studies\,\cite{Rosuel_arxiv,Kittaka2020}. Focusing on the initial slope, we find the anisotropy of $H_{\rm c2}^b>H_{\rm c2}^a>H_{\rm c2}^c$. In contrast, from the initial slope of $H_{\rm c1}$ data (see Fig.\,\ref{Fig3}), we obtain the anisotropy $H_{\rm c1}^c>H_{\rm c1}^b>H_{\rm c1}^a$, the order of which is not as expected from the anisotropy of $H_{\rm c2}$. We note that, while the anisotropy in $H_{\rm c1}$ is similar to the previous study of global magnetization measurements in a lower-quality sample\,\cite{Paulsen2021}, the absolute values of $H_{\rm c1}$ in this study are considerably larger. This point is important to discuss the origin of the discrepancy between the anisotropies of $H_{\rm c1}$ and $H_{\rm c2}$ as described below.

To discuss possible origins of this apparent inconsistency in critical field anisotropies, we calculate the lower critical field $H_{\rm c1}^{\rm cal}$ expected from $H_{\rm c}$ and $H_{\rm c2}$ values via the following GL relations,
\begin{eqnarray}
  H_{\rm c2} &=& \sqrt{2}\kappa H_{\rm c} \\
  H_{\rm c1}^{\rm cal} &=& \frac{H_{\rm c}}{\sqrt{2}\kappa} \left( \ln \kappa + 0.5 \right).
\end{eqnarray}
The results are shown in Fig.\,\ref{Fig3} as the black broken lines. We can clearly find in Fig.\,\ref{Fig3}(a) that $H_{\rm c1}^{\rm cal}$ along $a$-axis matches well with the experimental $H_{\rm c1}$ value, meaning that the experimental critical fields ($H_{\rm c}$, $H_{\rm c1}$, and $H_{\rm c2}$) along $a$-axis satisfy the usual thermodynamic relations of Eqs.\,(3,4). This result suggests that the significant change in the slope of $H_{\rm c2}^a$ near $T_{\rm c}$ shown in Fig.\,\ref{Fig4} is an intrinsic property of UTe$_2$ possibly induced by the reduction of ferromagnetic fluctuations with ${\bm H} \parallel {\bm a}$\,\cite{Rosuel_arxiv}. In the previous study, because the critical fields violated the above GL relations, the authors assumed a magnetic field effect on $T_{\rm c}$, $dT_{\rm c}/dH$\,\cite{Paulsen2021}. However, our results consistent with the GL relations for $H\parallel a$ indicate that the $dT_{\rm c}/dH$ effect is negligible as long as we focus on the initial slopes of the critical fields.

In contrast to $H_{\rm c1}^a$, however, the initial slope of $H_{\rm c1}^b$ is significantly larger than the expected $H_{\rm c1}^{\rm cal}$ value. In addition to the steep initial slope, $H_{\rm c1}^b$ and $H_{\rm c1}^c$ in ultraclean UTe$_2$ show an unusual increase below 0.5\,K, which has not been observed previously. A similar $H_{\rm c1}(T)$ behavior was reported in multi-band iron-based superconductors\,\cite{Song2011,Adamski2017} and the chiral superconductor candidates, UPt$_3$\,\cite{Vincent1991,Amann1998} and PrOs$_4$Sb$_{12}$\,\cite{Cichorek2005}. We emphasize that the increase of $H_{\rm c1}(T)$ at low temperatures in these materials was observed regardless of the magnetic field directions, which is clearly different from the case in UTe$_2$ where the increase is discernible only in ${\bm H} \parallel {\bm b}$ and ${\bm H} \parallel {\bm c}$. Moreover, magnetic penetration depth measurements in PrOs$_4$Sb$_{12}$ also show a kink at lower temperatures\,\cite{Chia2003}, while the anisotropic penetration depth in UTe$_2$ shows smooth $T$ dependence around 0.5\,K\,\cite{Ishihara_arxiv}. Thus, while the enhancement of $H_{\rm c1}$ at low temperatures in the previous studies on the various superconductors reflects the increase of superfluid density, the anomalous $H_{\rm c1}(T)$ in UTe$_2$ likely has a different origin which induces a large anisotropy.

Because the $H_{\rm c1}$ value is determined by the vortex line energy, the unusual $H_{\rm c1}^b(T)$ and $H_{\rm c1}^c(T)$ behaviors can be related to an exotic vortex state. As for the vortices in UTe$_2$, recent scanning SQUID measurements on the (011)-plane observed pinned vortices and antivortices even in a zero-field cooling condition, which can be related to the presence of strong and slow magnetic fluctuations detected in $\mu$SR and NMR studies\,\cite{Iguchi_arxiv}. Moreover, optical Kerr effect measurements on the (001)-plane suggest the presence of magnetized vortices induced by strong magnetic fluctuations. From these results, we consider that the enhancement of $H_{\rm c1}^b(T)$ and $H_{\rm c1}^c(T)$ is related to the strong Ising-like ferromagnetic fluctuations along $a$-axis inducing an exotic vortex state in UTe$_2$. We note that anomalous low-$T$ behavior has been also detected in $1/T_1 T$ of NMR measurements\,\cite{Nakamine2019} and zero-field relaxation rate of $\mu$SR measurements\,\cite{Sundar2019,Sundar_arxiv}.

The relationship between strong magnetic fluctuations and the enhancement in $H_{\rm c1}$ has been investigated in BaFe$_2$(As$_{1-x}$P$_x$)$_2$\,\cite{Putzke2014}. In this system, an antiferromagnetic quantum critical point (QCP) appears at $x=0.3$ where $\lambda(0)$ shows a sharp peak reflecting the strong mass enhancement\,\cite{Hashimoto2012}. In contrast, $H_{\rm c1}$ also shows a peak at QCP, contradicting with the expected behavior from Eq.\,(2). This discrepancy suggests that the vortex line energy is significantly enhanced by the strong magnetic fluctuations near QCP. Thus, considering the presence of strong anisotropic ferromagnetic fluctuations in UTe$_2$, the vortex line energy can be anisotropically enhanced by the Ising-like magnetic fluctuations. Indeed, the normal-state NMR measurements in UTe$_2$\,\cite{Tokunaga2019} reported that $1/T_1T$ shows a pronounced low-$T$ enhancement for ${\bm H} \parallel {\bm b}$ and ${\bm H} \parallel {\bm c}$, which is absent for ${\bm H} \parallel {\bm a}$. This shows a good correspondence with the anisotropic $H_{\rm c1}$ enhancement.

In conclusion, we measured the critical fields, $H_{\rm c}(T)$, $H_{\rm c1}(T)$, and $H_{\rm c2}(T)$, in high-quality single crystals of UTe$_2$ with $T_{\rm c}=2.1$\,K. While the three critical fields for ${\bm H} \parallel {\bm a}$ satisfy the usual GL relations near $T_{\rm c}$, the experimental $H_{\rm c1}^b$ and $H_{\rm c1}^c$ are larger than the expected values, which become more apparent below 0.5\,K. These results indicate that the anisotropic ferromagnetic fluctuations in UTe$_2$ significantly enhance the vortex line energy with ${\bm H} \parallel {\bm b}$ and ${\bm H} \parallel {\bm c}$. Our experimental results not only suggest that an exotic vortex state caused by anisotropic ferromagnetic fluctuations is realized in UTe$_2$, but also promote further studies on magnetic fluctuations in high-quality single crystals of UTe$_2$ with low magnetic impurity density.

We thank J.-P. Brison for fruitful discussions. This work was supported by Grants-in-Aid for Scientific Research (KAKENHI) (Nos. JP22H00105, JP22K20349, JP21H01793, JP19H00649, JP18H05227), Grant-in-Aid for Scientific Research on innovative areas ``Quantum Liquid Crystals'' (No. JP19H05824), Grant-in-Aid for Scientific Research for Transformative Research Areas (A) ``Condensed Conjugation'' (No. JP20H05869) from Japan Society for the Promotion of Science (JSPS), and CREST (No. JPMJCR19T5) from Japan Science and Technology (JST).

\clearpage

\renewcommand{\tablename}{Table S$\!\!$}
\renewcommand{\thetable}{\arabic{table}}
\renewcommand{\figurename}{Figure S$\!\!$}
\renewcommand*{\citenumfont}[1]{S#1}
\renewcommand*{\bibnumfmt}[1]{[S#1]}
\renewcommand{\theequation}{S\arabic{equation}}

\begin{center}
{\Large \bf Supplemental Material}
\end{center}

\section{Calculation of $H_{\rm c}$ from specific heat}

\begin{figure}[b]
    \includegraphics[width=\linewidth]{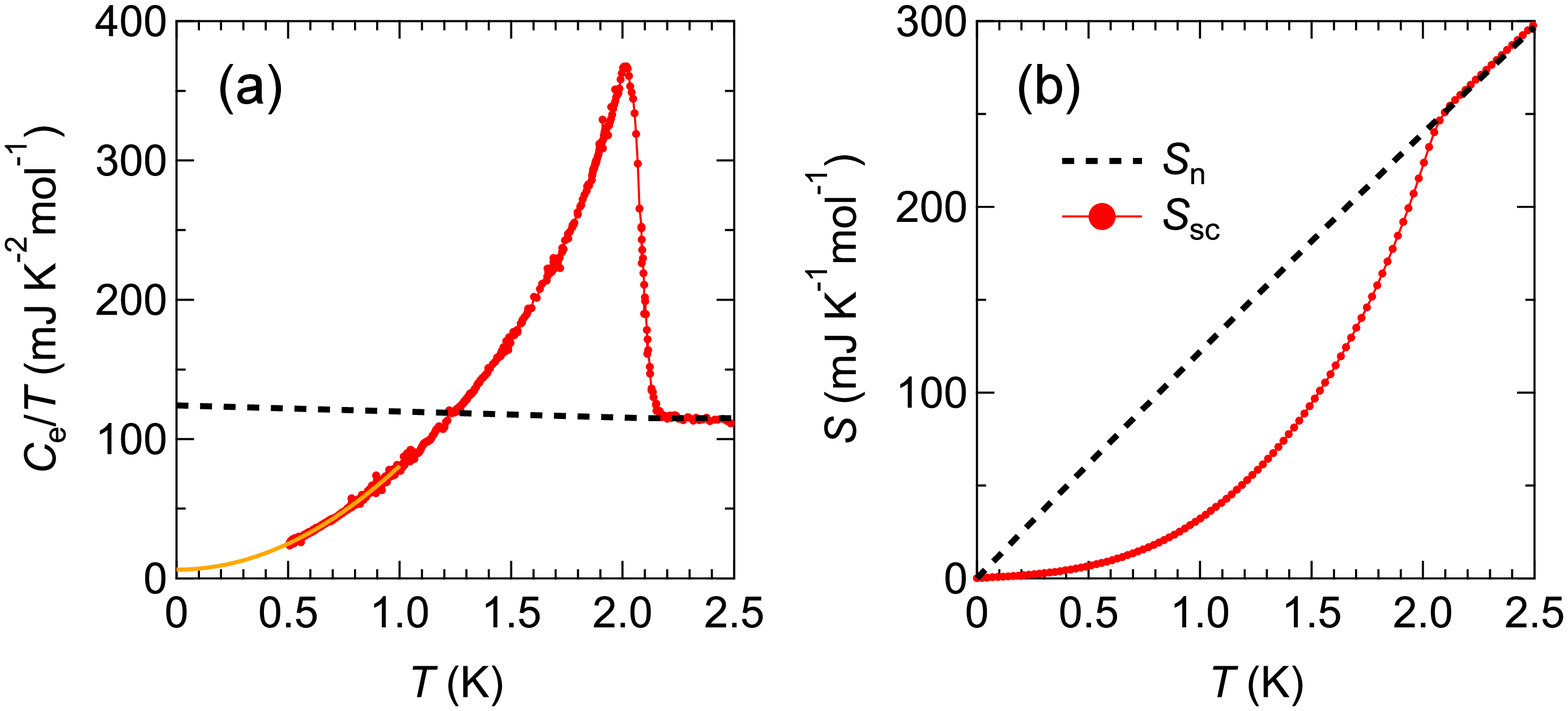}
    \caption{(a)\,Electronic specific heat $C_{\rm e}$ divided by temperature $T$ as a function of $T$ obtained by subtracting the $T^3$ component from the total specific heat. The orange line represents an extrapolation curve toward $T=0$. The black broken line is $C_{\rm e}/T$ in the normal state including a small linear term below $T_{\rm c}$. (b)\,Temperature dependence of entropy in the normal state (black broken line) and the superconducting state (red markers).}
    \label{FigS1}
\end{figure}

Figure\,S\ref{FigS1}(a) shows the electronic specific heat of crystal \#A1 evaluated by subtracting a $C\propto T^3$ component from the total specific heat in Fig.\,1(a). To calculate the entropy in the superconducting state $S_{\rm sc}$, we extrapolated the data toward $T=0$ using the fitting function $C_{\rm e}/T = a+bT^2$ in the low-$T$ region (the orange curve in Fig.\,S\ref{FigS1}(a)). The residual value $a=6.3$\,mJK$^{-2}$mol$^{-1}$ is much smaller than the reported values in the CVT crystals\,[2,23], confirming the ultraclean nature of our crystals. The $S_{\rm sc}$ values calculated from the extrapolated $C_{\rm e}/T$ data are the red markers in Fig.\,S\ref{FigS1}(b). Then, we need to evaluate the normal state entropy $S_{\rm n}$ to obtain $H_{\rm c}(T)$. In general, $S_{\rm sc}$ and $S_{\rm n}$ should be equal at $T_{\rm c}$ because the superconducting transition is of second order. However, we find that this entropy balance is violated assuming that $C_{\rm e}/T$ is completely constant in the normal state below $T_{\rm c}$. This result is caused by the small residual value at $T=0$ and an upturn behavior in the low-$T$ region observed in previous studies\,[36,45]. To satisfy the entropy balance, for simplicity, we introduced a small linear $T$ dependence below $T_{\rm c}$ in $C_{\rm e}/T$ of the normal state. We consider that, since this linear term is small compared with the $T$ dependence of $C_{\rm e}/T$ in the superconducting state, the $T$ dependence of $C_{\rm e}/T$ in the normal state does not affect much the estimation of $H_{\rm c}(T)$. Finally, we obtain $C_{\rm e}/T$ in the normal state and $S_{\rm n}$ shown in Fig.\,S\ref{FigS1}(a) and (b) as the black broken lines, respectively. As mentioned in the main text, we can calculate $H_{\rm c}$ through the equation, $H_{\rm c}^2(T)=2\mu_0 \int_T^{T_{\rm c}} (S_{\rm n}-S_{\rm sc}) dT$.

\section{Position dependence of the local magnetic induction}

\begin{figure}[b]
    \includegraphics[width=\linewidth]{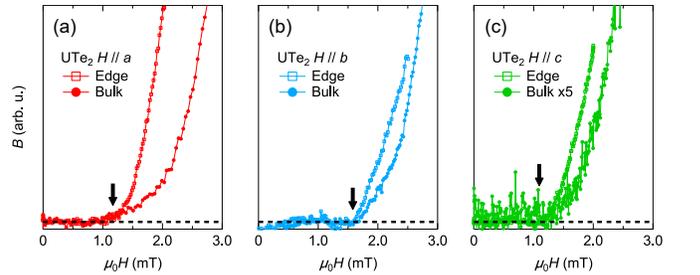}
    \caption{(a-c)\,Local magnetic induction near the crystal edges and in the bulk as a function of external magnetic field at 1\,K for ${\bm H} \parallel {\bm a}$, ${\bm H} \parallel {\bm b}$, and ${\bm H} \parallel {\bm c}$, respectively. The black arrows are $H_{\rm p}$ estimated from the edge data.}
    \label{FigS2}
\end{figure}

When the surface pinning or geometrical barriers of superconducting vortices are present, $H_{\rm p}$ is much enhanced from $H_{\rm c1}$ value and shows a large position dependence\,[31-33]. To consider this effect, we measured the position dependence of $H_{\rm p}$. Figure\,S\ref{FigS2}(a-c) shows local magnetic induction near edges (open squares) and in the bulk (filled circles) as a function of external magnetic field at 1\,K for ${\bm H} \parallel {\bm a}$, ${\bm H} \parallel {\bm b}$, and ${\bm H} \parallel {\bm c}$, respectively. Because the distance of neighboring Hall bars is 20\,$\mu$m, the bulk data reflect the magnetic induction away from the crystal edges by at least 20\,$\mu$m. Obviously, we find that the $H_{\rm p}$ values are almost independent on the positions. This result support that our measurements are not affected by the surface pinning or geometrical barriers of the superconducting vortices.

Regarding the Bean model, the magnetic induction $B$ satisfies the relation, $B\propto (H-H_{\rm p})^2$, above $H_{\rm p}$. Thus, we fitted $B^{1/2} (H)$ with a linear function in a $B>0$ region and defined $H_{\rm p}$ as the crossing point of the fitting line and $B=0$.


\begin{thebibliography}{46}%
\makeatletter
\providecommand \@ifxundefined [1]{%
 \@ifx{#1\undefined}
}%
\providecommand \@ifnum [1]{%
 \ifnum #1\expandafter \@firstoftwo
 \else \expandafter \@secondoftwo
 \fi
}%
\providecommand \@ifx [1]{%
 \ifx #1\expandafter \@firstoftwo
 \else \expandafter \@secondoftwo
 \fi
}%
\providecommand \natexlab [1]{#1}%
\providecommand \enquote  [1]{``#1''}%
\providecommand \bibnamefont  [1]{#1}%
\providecommand \bibfnamefont [1]{#1}%
\providecommand \citenamefont [1]{#1}%
\providecommand \href@noop [0]{\@secondoftwo}%
\providecommand \href [0]{\begingroup \@sanitize@url \@href}%
\providecommand \@href[1]{\@@startlink{#1}\@@href}%
\providecommand \@@href[1]{\endgroup#1\@@endlink}%
\providecommand \@sanitize@url [0]{\catcode `\\12\catcode `\$12\catcode
  `\&12\catcode `\#12\catcode `\^12\catcode `\_12\catcode `\%12\relax}%
\providecommand \@@startlink[1]{}%
\providecommand \@@endlink[0]{}%
\providecommand \url  [0]{\begingroup\@sanitize@url \@url }%
\providecommand \@url [1]{\endgroup\@href {#1}{\urlprefix }}%
\providecommand \urlprefix  [0]{URL }%
\providecommand \Eprint [0]{\href }%
\providecommand \doibase [0]{http://dx.doi.org/}%
\providecommand \selectlanguage [0]{\@gobble}%
\providecommand \bibinfo  [0]{\@secondoftwo}%
\providecommand \bibfield  [0]{\@secondoftwo}%
\providecommand \translation [1]{[#1]}%
\providecommand \BibitemOpen [0]{}%
\providecommand \bibitemStop [0]{}%
\providecommand \bibitemNoStop [0]{.\EOS\space}%
\providecommand \EOS [0]{\spacefactor3000\relax}%
\providecommand \BibitemShut  [1]{\csname bibitem#1\endcsname}%
\let\auto@bib@innerbib\@empty
\bibitem [{\citenamefont {Ran}\ \emph {et~al.}(2019{\natexlab{a}})\citenamefont
  {Ran}, \citenamefont {Eckberg}, \citenamefont {Ding}, \citenamefont
  {Furukawa}, \citenamefont {Metz}, \citenamefont {Saha}, \citenamefont {Liu},
  \citenamefont {Zic}, \citenamefont {Kim}, \citenamefont {Paglione},\ and\
  \citenamefont {Butch}}]{Ran2019S}%
  \BibitemOpen
  \bibfield  {author} {\bibinfo {author} {\bibfnamefont {S.}~\bibnamefont
  {Ran}}, \bibinfo {author} {\bibfnamefont {C.}~\bibnamefont {Eckberg}},
  \bibinfo {author} {\bibfnamefont {Q.-P.}\ \bibnamefont {Ding}}, \bibinfo
  {author} {\bibfnamefont {Y.}~\bibnamefont {Furukawa}}, \bibinfo {author}
  {\bibfnamefont {T.}~\bibnamefont {Metz}}, \bibinfo {author} {\bibfnamefont
  {S.~R.}\ \bibnamefont {Saha}}, \bibinfo {author} {\bibfnamefont {I.-L.}\
  \bibnamefont {Liu}}, \bibinfo {author} {\bibfnamefont {M.}~\bibnamefont
  {Zic}}, \bibinfo {author} {\bibfnamefont {H.}~\bibnamefont {Kim}}, \bibinfo
  {author} {\bibfnamefont {J.}~\bibnamefont {Paglione}}, \ and\ \bibinfo
  {author} {\bibfnamefont {N.~P.}\ \bibnamefont {Butch}},\ }\href {\doibase
  10.1126/science.aav8645} {\bibfield  {journal} {\bibinfo  {journal}
  {Science}\ }\textbf {\bibinfo {volume} {365}},\ \bibinfo {pages} {684}
  (\bibinfo {year} {2019}{\natexlab{a}})}\BibitemShut {NoStop}%
\bibitem [{\citenamefont {Aoki}\ \emph
  {et~al.}(2022{\natexlab{a}})\citenamefont {Aoki}, \citenamefont {Brison},
  \citenamefont {Flouquet}, \citenamefont {Ishida}, \citenamefont {Knebel},
  \citenamefont {Tokunaga},\ and\ \citenamefont {Yanase}}]{Aoki2022R}%
  \BibitemOpen
  \bibfield  {author} {\bibinfo {author} {\bibfnamefont {D.}~\bibnamefont
  {Aoki}}, \bibinfo {author} {\bibfnamefont {J.-P.}\ \bibnamefont {Brison}},
  \bibinfo {author} {\bibfnamefont {J.}~\bibnamefont {Flouquet}}, \bibinfo
  {author} {\bibfnamefont {K.}~\bibnamefont {Ishida}}, \bibinfo {author}
  {\bibfnamefont {G.}~\bibnamefont {Knebel}}, \bibinfo {author} {\bibfnamefont
  {Y.}~\bibnamefont {Tokunaga}}, \ and\ \bibinfo {author} {\bibfnamefont
  {Y.}~\bibnamefont {Yanase}},\ }\href {\doibase 10.1088/1361-648X/ac5863}
  {\bibfield  {journal} {\bibinfo  {journal} {J. Phys.: Condens. Matter}\
  }\textbf {\bibinfo {volume} {34}},\ \bibinfo {pages} {243002} (\bibinfo
  {year} {2022}{\natexlab{a}})}\BibitemShut {NoStop}%
\bibitem [{\citenamefont {Ran}\ \emph {et~al.}(2019{\natexlab{b}})\citenamefont
  {Ran}, \citenamefont {Liu}, \citenamefont {Eo}, \citenamefont {Campbell},
  \citenamefont {Neves}, \citenamefont {Fuhrman}, \citenamefont {Saha},
  \citenamefont {Eckberg}, \citenamefont {Kim}, \citenamefont {Graf},
  \citenamefont {Balakirev}, \citenamefont {Singleton}, \citenamefont
  {Paglione},\ and\ \citenamefont {Butch}}]{Ran2019NP}%
  \BibitemOpen
  \bibfield  {author} {\bibinfo {author} {\bibfnamefont {S.}~\bibnamefont
  {Ran}}, \bibinfo {author} {\bibfnamefont {I.-L.}\ \bibnamefont {Liu}},
  \bibinfo {author} {\bibfnamefont {Y.~S.}\ \bibnamefont {Eo}}, \bibinfo
  {author} {\bibfnamefont {D.~J.}\ \bibnamefont {Campbell}}, \bibinfo {author}
  {\bibfnamefont {P.~M.}\ \bibnamefont {Neves}}, \bibinfo {author}
  {\bibfnamefont {W.~T.}\ \bibnamefont {Fuhrman}}, \bibinfo {author}
  {\bibfnamefont {S.~R.}\ \bibnamefont {Saha}}, \bibinfo {author}
  {\bibfnamefont {C.}~\bibnamefont {Eckberg}}, \bibinfo {author} {\bibfnamefont
  {H.}~\bibnamefont {Kim}}, \bibinfo {author} {\bibfnamefont {D.}~\bibnamefont
  {Graf}}, \bibinfo {author} {\bibfnamefont {F.}~\bibnamefont {Balakirev}},
  \bibinfo {author} {\bibfnamefont {J.}~\bibnamefont {Singleton}}, \bibinfo
  {author} {\bibfnamefont {J.}~\bibnamefont {Paglione}}, \ and\ \bibinfo
  {author} {\bibfnamefont {N.~P.}\ \bibnamefont {Butch}},\ }\href {\doibase
  10.1038/s41567-019-0670-x} {\bibfield  {journal} {\bibinfo  {journal} {Nat.
  Phys.}\ }\textbf {\bibinfo {volume} {15}},\ \bibinfo {pages} {1250} (\bibinfo
  {year} {2019}{\natexlab{b}})}\BibitemShut {NoStop}%
\bibitem [{\citenamefont {Braithwaite}\ \emph {et~al.}(2019)\citenamefont
  {Braithwaite}, \citenamefont {Vali{\v s}ka}, \citenamefont {Knebel},
  \citenamefont {Lapertot}, \citenamefont {Brison}, \citenamefont {Pourret},
  \citenamefont {Zhitomirsky}, \citenamefont {Flouquet}, \citenamefont
  {Honda},\ and\ \citenamefont {Aoki}}]{Braithwaite2019}%
  \BibitemOpen
  \bibfield  {author} {\bibinfo {author} {\bibfnamefont {D.}~\bibnamefont
  {Braithwaite}}, \bibinfo {author} {\bibfnamefont {M.}~\bibnamefont {Vali{\v
  s}ka}}, \bibinfo {author} {\bibfnamefont {G.}~\bibnamefont {Knebel}},
  \bibinfo {author} {\bibfnamefont {G.}~\bibnamefont {Lapertot}}, \bibinfo
  {author} {\bibfnamefont {J.-P.}\ \bibnamefont {Brison}}, \bibinfo {author}
  {\bibfnamefont {A.}~\bibnamefont {Pourret}}, \bibinfo {author} {\bibfnamefont
  {M.~E.}\ \bibnamefont {Zhitomirsky}}, \bibinfo {author} {\bibfnamefont
  {J.}~\bibnamefont {Flouquet}}, \bibinfo {author} {\bibfnamefont
  {F.}~\bibnamefont {Honda}}, \ and\ \bibinfo {author} {\bibfnamefont
  {D.}~\bibnamefont {Aoki}},\ }\href {\doibase 10.1038/s42005-019-0248-z}
  {\bibfield  {journal} {\bibinfo  {journal} {Commun. Phys.}\ }\textbf
  {\bibinfo {volume} {2}},\ \bibinfo {pages} {147} (\bibinfo {year}
  {2019})}\BibitemShut {NoStop}%
\bibitem [{\citenamefont {Thomas}\ \emph {et~al.}(2020)\citenamefont {Thomas},
  \citenamefont {Santos}, \citenamefont {Christensen}, \citenamefont {Asaba},
  \citenamefont {Ronning}, \citenamefont {Thompson}, \citenamefont {Bauer},
  \citenamefont {Fernandes}, \citenamefont {Fabbris},\ and\ \citenamefont
  {Rosa}}]{Thomas2020}%
  \BibitemOpen
  \bibfield  {author} {\bibinfo {author} {\bibfnamefont {S.~M.}\ \bibnamefont
  {Thomas}}, \bibinfo {author} {\bibfnamefont {F.~B.}\ \bibnamefont {Santos}},
  \bibinfo {author} {\bibfnamefont {M.~H.}\ \bibnamefont {Christensen}},
  \bibinfo {author} {\bibfnamefont {T.}~\bibnamefont {Asaba}}, \bibinfo
  {author} {\bibfnamefont {F.}~\bibnamefont {Ronning}}, \bibinfo {author}
  {\bibfnamefont {J.~D.}\ \bibnamefont {Thompson}}, \bibinfo {author}
  {\bibfnamefont {E.~D.}\ \bibnamefont {Bauer}}, \bibinfo {author}
  {\bibfnamefont {R.~M.}\ \bibnamefont {Fernandes}}, \bibinfo {author}
  {\bibfnamefont {G.}~\bibnamefont {Fabbris}}, \ and\ \bibinfo {author}
  {\bibfnamefont {P.~F.~S.}\ \bibnamefont {Rosa}},\ }\href {\doibase
  10.1126/sciadv.abc8709} {\bibfield  {journal} {\bibinfo  {journal} {Sci.
  Adv.}\ }\textbf {\bibinfo {volume} {6}},\ \bibinfo {pages} {eabc8709}
  (\bibinfo {year} {2020})}\BibitemShut {NoStop}%
\bibitem [{\citenamefont {Aoki}\ \emph {et~al.}(2020)\citenamefont {Aoki},
  \citenamefont {Honda}, \citenamefont {Knebel}, \citenamefont {Braithwaite},
  \citenamefont {Nakamura}, \citenamefont {Li}, \citenamefont {Homma},
  \citenamefont {Shimizu}, \citenamefont {Sato}, \citenamefont {Brison},\ and\
  \citenamefont {Flouquet}}]{Aoki2020}%
  \BibitemOpen
  \bibfield  {author} {\bibinfo {author} {\bibfnamefont {D.}~\bibnamefont
  {Aoki}}, \bibinfo {author} {\bibfnamefont {F.}~\bibnamefont {Honda}},
  \bibinfo {author} {\bibfnamefont {G.}~\bibnamefont {Knebel}}, \bibinfo
  {author} {\bibfnamefont {D.}~\bibnamefont {Braithwaite}}, \bibinfo {author}
  {\bibfnamefont {A.}~\bibnamefont {Nakamura}}, \bibinfo {author}
  {\bibfnamefont {D.}~\bibnamefont {Li}}, \bibinfo {author} {\bibfnamefont
  {Y.}~\bibnamefont {Homma}}, \bibinfo {author} {\bibfnamefont
  {Y.}~\bibnamefont {Shimizu}}, \bibinfo {author} {\bibfnamefont {Y.~J.}\
  \bibnamefont {Sato}}, \bibinfo {author} {\bibfnamefont {J.-P.}\ \bibnamefont
  {Brison}}, \ and\ \bibinfo {author} {\bibfnamefont {J.}~\bibnamefont
  {Flouquet}},\ }\href {\doibase 10.7566/JPSJ.89.053705} {\bibfield  {journal}
  {\bibinfo  {journal} {J. Phys. Soc. Jpn.}\ }\textbf {\bibinfo {volume}
  {89}},\ \bibinfo {pages} {053705} (\bibinfo {year} {2020})}\BibitemShut
  {NoStop}%
\bibitem [{\citenamefont {Kinjo}\ \emph {et~al.}()\citenamefont {Kinjo},
  \citenamefont {Fujibayashi}, \citenamefont {Kitagawa}, \citenamefont
  {Ishida}, \citenamefont {Tokunaga}, \citenamefont {Sakai}, \citenamefont
  {Kambe}, \citenamefont {Nakamura}, \citenamefont {Shimizu}, \citenamefont
  {Homma}, \citenamefont {Li}, \citenamefont {Honda}, \citenamefont {Aoki},
  \citenamefont {Hiraki}, \citenamefont {Kimata},\ and\ \citenamefont
  {Sasaki}}]{Kinjo_arxiv}%
  \BibitemOpen
  \bibfield  {author} {\bibinfo {author} {\bibfnamefont {K.}~\bibnamefont
  {Kinjo}}, \bibinfo {author} {\bibfnamefont {H.}~\bibnamefont {Fujibayashi}},
  \bibinfo {author} {\bibfnamefont {S.}~\bibnamefont {Kitagawa}}, \bibinfo
  {author} {\bibfnamefont {K.}~\bibnamefont {Ishida}}, \bibinfo {author}
  {\bibfnamefont {Y.}~\bibnamefont {Tokunaga}}, \bibinfo {author}
  {\bibfnamefont {H.}~\bibnamefont {Sakai}}, \bibinfo {author} {\bibfnamefont
  {S.}~\bibnamefont {Kambe}}, \bibinfo {author} {\bibfnamefont
  {A.}~\bibnamefont {Nakamura}}, \bibinfo {author} {\bibfnamefont
  {Y.}~\bibnamefont {Shimizu}}, \bibinfo {author} {\bibfnamefont
  {Y.}~\bibnamefont {Homma}}, \bibinfo {author} {\bibfnamefont {D.~X.}\
  \bibnamefont {Li}}, \bibinfo {author} {\bibfnamefont {F.}~\bibnamefont
  {Honda}}, \bibinfo {author} {\bibfnamefont {D.}~\bibnamefont {Aoki}},
  \bibinfo {author} {\bibfnamefont {K.}~\bibnamefont {Hiraki}}, \bibinfo
  {author} {\bibfnamefont {M.}~\bibnamefont {Kimata}}, \ and\ \bibinfo {author}
  {\bibfnamefont {T.}~\bibnamefont {Sasaki}},\ }\href@noop {} {\bibinfo
  {journal} {arXiv:2206.02444}\ }\BibitemShut {NoStop}%
\bibitem [{\citenamefont {Rosuel}\ \emph {et~al.}()\citenamefont {Rosuel},
  \citenamefont {Marcenat}, \citenamefont {Knebel}, \citenamefont {Klein},
  \citenamefont {Pourret}, \citenamefont {Marquardt}, \citenamefont {Niu},
  \citenamefont {Rousseau}, \citenamefont {Demuer}, \citenamefont {Seyfarth},
  \citenamefont {Lapertot}, \citenamefont {Aoki}, \citenamefont {Braithwaite},
  \citenamefont {Flouquet},\ and\ \citenamefont {Brison}}]{Rosuel_arxiv}%
  \BibitemOpen
\bibfield  {journal} {  }\bibfield  {author} {\bibinfo {author} {\bibfnamefont
  {A.}~\bibnamefont {Rosuel}}, \bibinfo {author} {\bibfnamefont
  {C.}~\bibnamefont {Marcenat}}, \bibinfo {author} {\bibfnamefont
  {G.}~\bibnamefont {Knebel}}, \bibinfo {author} {\bibfnamefont
  {T.}~\bibnamefont {Klein}}, \bibinfo {author} {\bibfnamefont
  {A.}~\bibnamefont {Pourret}}, \bibinfo {author} {\bibfnamefont
  {N.}~\bibnamefont {Marquardt}}, \bibinfo {author} {\bibfnamefont
  {Q.}~\bibnamefont {Niu}}, \bibinfo {author} {\bibfnamefont {S.}~\bibnamefont
  {Rousseau}}, \bibinfo {author} {\bibfnamefont {A.}~\bibnamefont {Demuer}},
  \bibinfo {author} {\bibfnamefont {G.}~\bibnamefont {Seyfarth}}, \bibinfo
  {author} {\bibfnamefont {G.}~\bibnamefont {Lapertot}}, \bibinfo {author}
  {\bibfnamefont {D.}~\bibnamefont {Aoki}}, \bibinfo {author} {\bibfnamefont
  {D.}~\bibnamefont {Braithwaite}}, \bibinfo {author} {\bibfnamefont
  {J.}~\bibnamefont {Flouquet}}, \ and\ \bibinfo {author} {\bibfnamefont
  {J.-P.}\ \bibnamefont {Brison}},\ }\href@noop {} {\bibinfo  {journal}
  {arXiv:2205.04524}\ }\BibitemShut {NoStop}%
\bibitem [{\citenamefont {Sakai}\ \emph {et~al.}()\citenamefont {Sakai},
  \citenamefont {Tokiwa}, \citenamefont {Opletal}, \citenamefont {Kimata},
  \citenamefont {Awaji}, \citenamefont {Sasaki}, \citenamefont {Aoki},
  \citenamefont {Kambe}, \citenamefont {Tokunaga},\ and\ \citenamefont
  {Haga}}]{Sakai_arxiv}%
  \BibitemOpen
\bibfield  {journal} {  }\bibfield  {author} {\bibinfo {author} {\bibfnamefont
  {H.}~\bibnamefont {Sakai}}, \bibinfo {author} {\bibfnamefont
  {Y.}~\bibnamefont {Tokiwa}}, \bibinfo {author} {\bibfnamefont
  {P.}~\bibnamefont {Opletal}}, \bibinfo {author} {\bibfnamefont
  {M.}~\bibnamefont {Kimata}}, \bibinfo {author} {\bibfnamefont
  {S.}~\bibnamefont {Awaji}}, \bibinfo {author} {\bibfnamefont
  {T.}~\bibnamefont {Sasaki}}, \bibinfo {author} {\bibfnamefont
  {D.}~\bibnamefont {Aoki}}, \bibinfo {author} {\bibfnamefont {S.}~\bibnamefont
  {Kambe}}, \bibinfo {author} {\bibfnamefont {Y.}~\bibnamefont {Tokunaga}}, \
  and\ \bibinfo {author} {\bibfnamefont {Y.}~\bibnamefont {Haga}},\ }\href@noop
  {} {\bibinfo  {journal} {arXiv:2210.05909}\ }\BibitemShut {NoStop}%
\bibitem [{\citenamefont {Jiao}\ \emph {et~al.}(2020)\citenamefont {Jiao},
  \citenamefont {Howard}, \citenamefont {Ran}, \citenamefont {Wang},
  \citenamefont {Rodriguez}, \citenamefont {Sigrist}, \citenamefont {Wang},
  \citenamefont {Butch},\ and\ \citenamefont {Madhavan}}]{Jiao2020}%
  \BibitemOpen
\bibfield  {journal} {  }\bibfield  {author} {\bibinfo {author} {\bibfnamefont
  {L.}~\bibnamefont {Jiao}}, \bibinfo {author} {\bibfnamefont {S.}~\bibnamefont
  {Howard}}, \bibinfo {author} {\bibfnamefont {S.}~\bibnamefont {Ran}},
  \bibinfo {author} {\bibfnamefont {Z.}~\bibnamefont {Wang}}, \bibinfo {author}
  {\bibfnamefont {J.~O.}\ \bibnamefont {Rodriguez}}, \bibinfo {author}
  {\bibfnamefont {M.}~\bibnamefont {Sigrist}}, \bibinfo {author} {\bibfnamefont
  {Z.}~\bibnamefont {Wang}}, \bibinfo {author} {\bibfnamefont {N.~P.}\
  \bibnamefont {Butch}}, \ and\ \bibinfo {author} {\bibfnamefont
  {V.}~\bibnamefont {Madhavan}},\ }\href {\doibase 10.1038/s41586-020-2122-2}
  {\bibfield  {journal} {\bibinfo  {journal} {Nature}\ }\textbf {\bibinfo
  {volume} {579}},\ \bibinfo {pages} {523} (\bibinfo {year}
  {2020})}\BibitemShut {NoStop}%
\bibitem [{\citenamefont {Hayes}\ \emph {et~al.}(2021)\citenamefont {Hayes},
  \citenamefont {Wei}, \citenamefont {Metz}, \citenamefont {Zhang},
  \citenamefont {Eo}, \citenamefont {Ran}, \citenamefont {Saha}, \citenamefont
  {Collini}, \citenamefont {Butch}, \citenamefont {Agterberg}, \citenamefont
  {Kapitulnik},\ and\ \citenamefont {Paglione}}]{Hayes2021}%
  \BibitemOpen
  \bibfield  {author} {\bibinfo {author} {\bibfnamefont {I.~M.}\ \bibnamefont
  {Hayes}}, \bibinfo {author} {\bibfnamefont {D.~S.}\ \bibnamefont {Wei}},
  \bibinfo {author} {\bibfnamefont {T.}~\bibnamefont {Metz}}, \bibinfo {author}
  {\bibfnamefont {J.}~\bibnamefont {Zhang}}, \bibinfo {author} {\bibfnamefont
  {Y.~S.}\ \bibnamefont {Eo}}, \bibinfo {author} {\bibfnamefont
  {S.}~\bibnamefont {Ran}}, \bibinfo {author} {\bibfnamefont {S.~R.}\
  \bibnamefont {Saha}}, \bibinfo {author} {\bibfnamefont {J.}~\bibnamefont
  {Collini}}, \bibinfo {author} {\bibfnamefont {N.~P.}\ \bibnamefont {Butch}},
  \bibinfo {author} {\bibfnamefont {D.~F.}\ \bibnamefont {Agterberg}}, \bibinfo
  {author} {\bibfnamefont {A.}~\bibnamefont {Kapitulnik}}, \ and\ \bibinfo
  {author} {\bibfnamefont {J.}~\bibnamefont {Paglione}},\ }\href {\doibase
  10.1126/science.abb0272} {\bibfield  {journal} {\bibinfo  {journal}
  {Science}\ }\textbf {\bibinfo {volume} {373}},\ \bibinfo {pages} {797}
  (\bibinfo {year} {2021})}\BibitemShut {NoStop}%
\bibitem [{\citenamefont {Bae}\ \emph {et~al.}(2021)\citenamefont {Bae},
  \citenamefont {Kim}, \citenamefont {Eo}, \citenamefont {Ran}, \citenamefont
  {Liu}, \citenamefont {Fuhrman}, \citenamefont {Paglione}, \citenamefont
  {Butch},\ and\ \citenamefont {Anlage}}]{Bae2021}%
  \BibitemOpen
  \bibfield  {author} {\bibinfo {author} {\bibfnamefont {S.}~\bibnamefont
  {Bae}}, \bibinfo {author} {\bibfnamefont {H.}~\bibnamefont {Kim}}, \bibinfo
  {author} {\bibfnamefont {Y.~S.}\ \bibnamefont {Eo}}, \bibinfo {author}
  {\bibfnamefont {S.}~\bibnamefont {Ran}}, \bibinfo {author} {\bibfnamefont
  {I.-l.}\ \bibnamefont {Liu}}, \bibinfo {author} {\bibfnamefont {W.~T.}\
  \bibnamefont {Fuhrman}}, \bibinfo {author} {\bibfnamefont {J.}~\bibnamefont
  {Paglione}}, \bibinfo {author} {\bibfnamefont {N.~P.}\ \bibnamefont {Butch}},
  \ and\ \bibinfo {author} {\bibfnamefont {S.~M.}\ \bibnamefont {Anlage}},\
  }\href {\doibase 10.1038/s41467-021-22906-6} {\bibfield  {journal} {\bibinfo
  {journal} {Nat. Commun.}\ }\textbf {\bibinfo {volume} {12}},\ \bibinfo
  {pages} {2644} (\bibinfo {year} {2021})}\BibitemShut {NoStop}%
\bibitem [{\citenamefont {Ishihara}\ \emph {et~al.}()\citenamefont {Ishihara},
  \citenamefont {Roppongi}, \citenamefont {Kobayashi}, \citenamefont
  {Mizukami}, \citenamefont {Sakai}, \citenamefont {Haga}, \citenamefont
  {Hashimoto},\ and\ \citenamefont {Shibauchi}}]{Ishihara_arxiv}%
  \BibitemOpen
  \bibfield  {author} {\bibinfo {author} {\bibfnamefont {K.}~\bibnamefont
  {Ishihara}}, \bibinfo {author} {\bibfnamefont {M.}~\bibnamefont {Roppongi}},
  \bibinfo {author} {\bibfnamefont {M.}~\bibnamefont {Kobayashi}}, \bibinfo
  {author} {\bibfnamefont {Y.}~\bibnamefont {Mizukami}}, \bibinfo {author}
  {\bibfnamefont {H.}~\bibnamefont {Sakai}}, \bibinfo {author} {\bibfnamefont
  {Y.}~\bibnamefont {Haga}}, \bibinfo {author} {\bibfnamefont {K.}~\bibnamefont
  {Hashimoto}}, \ and\ \bibinfo {author} {\bibfnamefont {T.}~\bibnamefont
  {Shibauchi}},\ }\href@noop {} {\bibinfo  {journal} {arXiv:2105.13721}\
  }\BibitemShut {NoStop}%
\bibitem [{\citenamefont {Knebel}\ \emph {et~al.}(2019)\citenamefont {Knebel},
  \citenamefont {Knafo}, \citenamefont {Pourret}, \citenamefont {Niu},
  \citenamefont {Vali\v{s}ka}, \citenamefont {Braithwaite}, \citenamefont
  {Lapertot}, \citenamefont {Nardone}, \citenamefont {Zitouni}, \citenamefont
  {Mishra}, \citenamefont {Sheikin}, \citenamefont {Seyfarth}, \citenamefont
  {Brison}, \citenamefont {Aoki},\ and\ \citenamefont {Flouquet}}]{Knebel2019}%
  \BibitemOpen
\bibfield  {journal} {  }\bibfield  {author} {\bibinfo {author} {\bibfnamefont
  {G.}~\bibnamefont {Knebel}}, \bibinfo {author} {\bibfnamefont
  {W.}~\bibnamefont {Knafo}}, \bibinfo {author} {\bibfnamefont
  {A.}~\bibnamefont {Pourret}}, \bibinfo {author} {\bibfnamefont
  {Q.}~\bibnamefont {Niu}}, \bibinfo {author} {\bibfnamefont {M.}~\bibnamefont
  {Vali\v{s}ka}}, \bibinfo {author} {\bibfnamefont {D.}~\bibnamefont
  {Braithwaite}}, \bibinfo {author} {\bibfnamefont {G.}~\bibnamefont
  {Lapertot}}, \bibinfo {author} {\bibfnamefont {M.}~\bibnamefont {Nardone}},
  \bibinfo {author} {\bibfnamefont {A.}~\bibnamefont {Zitouni}}, \bibinfo
  {author} {\bibfnamefont {S.}~\bibnamefont {Mishra}}, \bibinfo {author}
  {\bibfnamefont {I.}~\bibnamefont {Sheikin}}, \bibinfo {author} {\bibfnamefont
  {G.}~\bibnamefont {Seyfarth}}, \bibinfo {author} {\bibfnamefont {J.-P.}\
  \bibnamefont {Brison}}, \bibinfo {author} {\bibfnamefont {D.}~\bibnamefont
  {Aoki}}, \ and\ \bibinfo {author} {\bibfnamefont {J.}~\bibnamefont
  {Flouquet}},\ }\href {\doibase 10.7566/JPSJ.88.063707} {\bibfield  {journal}
  {\bibinfo  {journal} {J. Phys. Soc. Jpn.}\ }\textbf {\bibinfo {volume}
  {88}},\ \bibinfo {pages} {063707} (\bibinfo {year} {2019})}\BibitemShut
  {NoStop}%
\bibitem [{\citenamefont {Aoki}\ \emph {et~al.}(2019)\citenamefont {Aoki},
  \citenamefont {Ishida},\ and\ \citenamefont {Flouquet}}]{Aoki2019}%
  \BibitemOpen
  \bibfield  {author} {\bibinfo {author} {\bibfnamefont {D.}~\bibnamefont
  {Aoki}}, \bibinfo {author} {\bibfnamefont {K.}~\bibnamefont {Ishida}}, \ and\
  \bibinfo {author} {\bibfnamefont {J.}~\bibnamefont {Flouquet}},\ }\href
  {\doibase 10.7566/JPSJ.88.022001} {\bibfield  {journal} {\bibinfo  {journal}
  {J. Phys. Soc. Jpn.}\ }\textbf {\bibinfo {volume} {88}},\ \bibinfo {pages}
  {022001} (\bibinfo {year} {2019})}\BibitemShut {NoStop}%
\bibitem [{\citenamefont {Tokunaga}\ \emph {et~al.}(2019)\citenamefont
  {Tokunaga}, \citenamefont {Sakai}, \citenamefont {Kambe}, \citenamefont
  {Hattori}, \citenamefont {Higa}, \citenamefont {Nakamine}, \citenamefont
  {Kitagawa}, \citenamefont {Ishida}, \citenamefont {Nakamura}, \citenamefont
  {Shimizu}, \citenamefont {Homma}, \citenamefont {Li}, \citenamefont {Honda},\
  and\ \citenamefont {Aoki}}]{Tokunaga2019}%
  \BibitemOpen
  \bibfield  {author} {\bibinfo {author} {\bibfnamefont {Y.}~\bibnamefont
  {Tokunaga}}, \bibinfo {author} {\bibfnamefont {H.}~\bibnamefont {Sakai}},
  \bibinfo {author} {\bibfnamefont {S.}~\bibnamefont {Kambe}}, \bibinfo
  {author} {\bibfnamefont {T.}~\bibnamefont {Hattori}}, \bibinfo {author}
  {\bibfnamefont {N.}~\bibnamefont {Higa}}, \bibinfo {author} {\bibfnamefont
  {G.}~\bibnamefont {Nakamine}}, \bibinfo {author} {\bibfnamefont
  {S.}~\bibnamefont {Kitagawa}}, \bibinfo {author} {\bibfnamefont
  {K.}~\bibnamefont {Ishida}}, \bibinfo {author} {\bibfnamefont
  {A.}~\bibnamefont {Nakamura}}, \bibinfo {author} {\bibfnamefont
  {Y.}~\bibnamefont {Shimizu}}, \bibinfo {author} {\bibfnamefont
  {Y.}~\bibnamefont {Homma}}, \bibinfo {author} {\bibfnamefont
  {D.}~\bibnamefont {Li}}, \bibinfo {author} {\bibfnamefont {F.}~\bibnamefont
  {Honda}}, \ and\ \bibinfo {author} {\bibfnamefont {D.}~\bibnamefont {Aoki}},\
  }\href {\doibase 10.7566/JPSJ.88.073701} {\bibfield  {journal} {\bibinfo
  {journal} {J. Phys. Soc. Jpn.}\ }\textbf {\bibinfo {volume} {88}},\ \bibinfo
  {pages} {073701} (\bibinfo {year} {2019})}\BibitemShut {NoStop}%
\bibitem [{\citenamefont {Sundar}\ \emph {et~al.}(2019)\citenamefont {Sundar},
  \citenamefont {Gheidi}, \citenamefont {Akintola}, \citenamefont {C\^ot\'e},
  \citenamefont {Dunsiger}, \citenamefont {Ran}, \citenamefont {Butch},
  \citenamefont {Saha}, \citenamefont {Paglione},\ and\ \citenamefont
  {Sonier}}]{Sundar2019}%
  \BibitemOpen
  \bibfield  {author} {\bibinfo {author} {\bibfnamefont {S.}~\bibnamefont
  {Sundar}}, \bibinfo {author} {\bibfnamefont {S.}~\bibnamefont {Gheidi}},
  \bibinfo {author} {\bibfnamefont {K.}~\bibnamefont {Akintola}}, \bibinfo
  {author} {\bibfnamefont {A.~M.}\ \bibnamefont {C\^ot\'e}}, \bibinfo {author}
  {\bibfnamefont {S.~R.}\ \bibnamefont {Dunsiger}}, \bibinfo {author}
  {\bibfnamefont {S.}~\bibnamefont {Ran}}, \bibinfo {author} {\bibfnamefont
  {N.~P.}\ \bibnamefont {Butch}}, \bibinfo {author} {\bibfnamefont {S.~R.}\
  \bibnamefont {Saha}}, \bibinfo {author} {\bibfnamefont {J.}~\bibnamefont
  {Paglione}}, \ and\ \bibinfo {author} {\bibfnamefont {J.~E.}\ \bibnamefont
  {Sonier}},\ }\href {\doibase 10.1103/PhysRevB.100.140502} {\bibfield
  {journal} {\bibinfo  {journal} {Phys. Rev. B}\ }\textbf {\bibinfo {volume}
  {100}},\ \bibinfo {pages} {140502(R)} (\bibinfo {year} {2019})}\BibitemShut
  {NoStop}%
\bibitem [{\citenamefont {Tokunaga}\ \emph {et~al.}(2022)\citenamefont
  {Tokunaga}, \citenamefont {Sakai}, \citenamefont {Kambe}, \citenamefont
  {Haga}, \citenamefont {Tokiwa}, \citenamefont {Opletal}, \citenamefont
  {Fujibayashi}, \citenamefont {Kinjo}, \citenamefont {Kitagawa}, \citenamefont
  {Ishida}, \citenamefont {Nakamura}, \citenamefont {Shimizu}, \citenamefont
  {Homma}, \citenamefont {Li}, \citenamefont {Honda},\ and\ \citenamefont
  {Aoki}}]{Tokunaga2022}%
  \BibitemOpen
  \bibfield  {author} {\bibinfo {author} {\bibfnamefont {Y.}~\bibnamefont
  {Tokunaga}}, \bibinfo {author} {\bibfnamefont {H.}~\bibnamefont {Sakai}},
  \bibinfo {author} {\bibfnamefont {S.}~\bibnamefont {Kambe}}, \bibinfo
  {author} {\bibfnamefont {Y.}~\bibnamefont {Haga}}, \bibinfo {author}
  {\bibfnamefont {Y.}~\bibnamefont {Tokiwa}}, \bibinfo {author} {\bibfnamefont
  {P.}~\bibnamefont {Opletal}}, \bibinfo {author} {\bibfnamefont
  {H.}~\bibnamefont {Fujibayashi}}, \bibinfo {author} {\bibfnamefont
  {K.}~\bibnamefont {Kinjo}}, \bibinfo {author} {\bibfnamefont
  {S.}~\bibnamefont {Kitagawa}}, \bibinfo {author} {\bibfnamefont
  {K.}~\bibnamefont {Ishida}}, \bibinfo {author} {\bibfnamefont
  {A.}~\bibnamefont {Nakamura}}, \bibinfo {author} {\bibfnamefont
  {Y.}~\bibnamefont {Shimizu}}, \bibinfo {author} {\bibfnamefont
  {Y.}~\bibnamefont {Homma}}, \bibinfo {author} {\bibfnamefont
  {D.}~\bibnamefont {Li}}, \bibinfo {author} {\bibfnamefont {F.}~\bibnamefont
  {Honda}}, \ and\ \bibinfo {author} {\bibfnamefont {D.}~\bibnamefont {Aoki}},\
  }\href {\doibase 10.7566/JPSJ.91.023707} {\bibfield  {journal} {\bibinfo
  {journal} {J. Phys. Soc. Jpn.}\ }\textbf {\bibinfo {volume} {91}},\ \bibinfo
  {pages} {023707} (\bibinfo {year} {2022})}\BibitemShut {NoStop}%
\bibitem [{\citenamefont {Duan}\ \emph {et~al.}(2020)\citenamefont {Duan},
  \citenamefont {Sasmal}, \citenamefont {Maple}, \citenamefont {Podlesnyak},
  \citenamefont {Zhu}, \citenamefont {Si},\ and\ \citenamefont
  {Dai}}]{Duan2020}%
  \BibitemOpen
  \bibfield  {author} {\bibinfo {author} {\bibfnamefont {C.}~\bibnamefont
  {Duan}}, \bibinfo {author} {\bibfnamefont {K.}~\bibnamefont {Sasmal}},
  \bibinfo {author} {\bibfnamefont {M.~B.}\ \bibnamefont {Maple}}, \bibinfo
  {author} {\bibfnamefont {A.}~\bibnamefont {Podlesnyak}}, \bibinfo {author}
  {\bibfnamefont {J.-X.}\ \bibnamefont {Zhu}}, \bibinfo {author} {\bibfnamefont
  {Q.}~\bibnamefont {Si}}, \ and\ \bibinfo {author} {\bibfnamefont
  {P.}~\bibnamefont {Dai}},\ }\href {\doibase 10.1103/PhysRevLett.125.237003}
  {\bibfield  {journal} {\bibinfo  {journal} {Phys. Rev. Lett.}\ }\textbf
  {\bibinfo {volume} {125}},\ \bibinfo {pages} {237003} (\bibinfo {year}
  {2020})}\BibitemShut {NoStop}%
\bibitem [{\citenamefont {Knafo}\ \emph
  {et~al.}(2021{\natexlab{a}})\citenamefont {Knafo}, \citenamefont {Knebel},
  \citenamefont {Steffens}, \citenamefont {Kaneko}, \citenamefont {Rosuel},
  \citenamefont {Brison}, \citenamefont {Flouquet}, \citenamefont {Aoki},
  \citenamefont {Lapertot},\ and\ \citenamefont {Raymond}}]{Knafo2021PRB}%
  \BibitemOpen
  \bibfield  {author} {\bibinfo {author} {\bibfnamefont {W.}~\bibnamefont
  {Knafo}}, \bibinfo {author} {\bibfnamefont {G.}~\bibnamefont {Knebel}},
  \bibinfo {author} {\bibfnamefont {P.}~\bibnamefont {Steffens}}, \bibinfo
  {author} {\bibfnamefont {K.}~\bibnamefont {Kaneko}}, \bibinfo {author}
  {\bibfnamefont {A.}~\bibnamefont {Rosuel}}, \bibinfo {author} {\bibfnamefont
  {J.-P.}\ \bibnamefont {Brison}}, \bibinfo {author} {\bibfnamefont
  {J.}~\bibnamefont {Flouquet}}, \bibinfo {author} {\bibfnamefont
  {D.}~\bibnamefont {Aoki}}, \bibinfo {author} {\bibfnamefont {G.}~\bibnamefont
  {Lapertot}}, \ and\ \bibinfo {author} {\bibfnamefont {S.}~\bibnamefont
  {Raymond}},\ }\href {\doibase 10.1103/PhysRevB.104.L100409} {\bibfield
  {journal} {\bibinfo  {journal} {Phys. Rev. B}\ }\textbf {\bibinfo {volume}
  {104}},\ \bibinfo {pages} {L100409} (\bibinfo {year}
  {2021}{\natexlab{a}})}\BibitemShut {NoStop}%
\bibitem [{\citenamefont {Duan}\ \emph {et~al.}(2021)\citenamefont {Duan},
  \citenamefont {Baumbach}, \citenamefont {Podlesnyak}, \citenamefont {Deng},
  \citenamefont {Moir}, \citenamefont {Breindel}, \citenamefont {Maple},
  \citenamefont {Nica}, \citenamefont {Si},\ and\ \citenamefont
  {Dai}}]{Duan2021}%
  \BibitemOpen
  \bibfield  {author} {\bibinfo {author} {\bibfnamefont {C.}~\bibnamefont
  {Duan}}, \bibinfo {author} {\bibfnamefont {R.~E.}\ \bibnamefont {Baumbach}},
  \bibinfo {author} {\bibfnamefont {A.}~\bibnamefont {Podlesnyak}}, \bibinfo
  {author} {\bibfnamefont {Y.}~\bibnamefont {Deng}}, \bibinfo {author}
  {\bibfnamefont {C.}~\bibnamefont {Moir}}, \bibinfo {author} {\bibfnamefont
  {A.~J.}\ \bibnamefont {Breindel}}, \bibinfo {author} {\bibfnamefont {M.~B.}\
  \bibnamefont {Maple}}, \bibinfo {author} {\bibfnamefont {E.~M.}\ \bibnamefont
  {Nica}}, \bibinfo {author} {\bibfnamefont {Q.}~\bibnamefont {Si}}, \ and\
  \bibinfo {author} {\bibfnamefont {P.}~\bibnamefont {Dai}},\ }\href {\doibase
  10.1038/s41586-021-04151-5} {\bibfield  {journal} {\bibinfo  {journal}
  {Nature}\ }\textbf {\bibinfo {volume} {600}},\ \bibinfo {pages} {636}
  (\bibinfo {year} {2021})}\BibitemShut {NoStop}%
\bibitem [{\citenamefont {Thomas}\ \emph {et~al.}(2021)\citenamefont {Thomas},
  \citenamefont {Stevens}, \citenamefont {Santos}, \citenamefont {Fender},
  \citenamefont {Bauer}, \citenamefont {Ronning}, \citenamefont {Thompson},
  \citenamefont {Huxley},\ and\ \citenamefont {Rosa}}]{Thomas2021}%
  \BibitemOpen
  \bibfield  {author} {\bibinfo {author} {\bibfnamefont {S.~M.}\ \bibnamefont
  {Thomas}}, \bibinfo {author} {\bibfnamefont {C.}~\bibnamefont {Stevens}},
  \bibinfo {author} {\bibfnamefont {F.~B.}\ \bibnamefont {Santos}}, \bibinfo
  {author} {\bibfnamefont {S.~S.}\ \bibnamefont {Fender}}, \bibinfo {author}
  {\bibfnamefont {E.~D.}\ \bibnamefont {Bauer}}, \bibinfo {author}
  {\bibfnamefont {F.}~\bibnamefont {Ronning}}, \bibinfo {author} {\bibfnamefont
  {J.~D.}\ \bibnamefont {Thompson}}, \bibinfo {author} {\bibfnamefont
  {A.}~\bibnamefont {Huxley}}, \ and\ \bibinfo {author} {\bibfnamefont
  {P.~F.~S.}\ \bibnamefont {Rosa}},\ }\href {\doibase
  10.1103/PhysRevB.104.224501} {\bibfield  {journal} {\bibinfo  {journal}
  {Phys. Rev. B}\ }\textbf {\bibinfo {volume} {104}},\ \bibinfo {pages}
  {224501} (\bibinfo {year} {2021})}\BibitemShut {NoStop}%
\bibitem [{\citenamefont {Rosa}\ \emph {et~al.}(2022)\citenamefont {Rosa},
  \citenamefont {Weiland}, \citenamefont {Fender}, \citenamefont {Scott},
  \citenamefont {Ronning}, \citenamefont {Thompson}, \citenamefont {Bauer},\
  and\ \citenamefont {Thomas}}]{Rosa2022}%
  \BibitemOpen
  \bibfield  {author} {\bibinfo {author} {\bibfnamefont {P.~F.~S.}\
  \bibnamefont {Rosa}}, \bibinfo {author} {\bibfnamefont {A.}~\bibnamefont
  {Weiland}}, \bibinfo {author} {\bibfnamefont {S.~S.}\ \bibnamefont {Fender}},
  \bibinfo {author} {\bibfnamefont {B.~L.}\ \bibnamefont {Scott}}, \bibinfo
  {author} {\bibfnamefont {F.}~\bibnamefont {Ronning}}, \bibinfo {author}
  {\bibfnamefont {J.~D.}\ \bibnamefont {Thompson}}, \bibinfo {author}
  {\bibfnamefont {E.~D.}\ \bibnamefont {Bauer}}, \ and\ \bibinfo {author}
  {\bibfnamefont {S.~M.}\ \bibnamefont {Thomas}},\ }\href {\doibase
  10.1038/s43246-022-00254-2} {\bibfield  {journal} {\bibinfo  {journal}
  {Commun. Mater.}\ }\textbf {\bibinfo {volume} {3}},\ \bibinfo {pages} {33}
  (\bibinfo {year} {2022})}\BibitemShut {NoStop}%
\bibitem [{\citenamefont {Knafo}\ \emph
  {et~al.}(2021{\natexlab{b}})\citenamefont {Knafo}, \citenamefont {Nardone},
  \citenamefont {Vali{\v s}ka}, \citenamefont {Zitouni}, \citenamefont
  {Lapertot}, \citenamefont {Aoki}, \citenamefont {Knebel},\ and\ \citenamefont
  {Braithwaite}}]{Knafo2021CP}%
  \BibitemOpen
  \bibfield  {author} {\bibinfo {author} {\bibfnamefont {W.}~\bibnamefont
  {Knafo}}, \bibinfo {author} {\bibfnamefont {M.}~\bibnamefont {Nardone}},
  \bibinfo {author} {\bibfnamefont {M.}~\bibnamefont {Vali{\v s}ka}}, \bibinfo
  {author} {\bibfnamefont {A.}~\bibnamefont {Zitouni}}, \bibinfo {author}
  {\bibfnamefont {G.}~\bibnamefont {Lapertot}}, \bibinfo {author}
  {\bibfnamefont {D.}~\bibnamefont {Aoki}}, \bibinfo {author} {\bibfnamefont
  {G.}~\bibnamefont {Knebel}}, \ and\ \bibinfo {author} {\bibfnamefont
  {D.}~\bibnamefont {Braithwaite}},\ }\href {\doibase
  10.1038/s42005-021-00545-z} {\bibfield  {journal} {\bibinfo  {journal}
  {Commun. Phys.}\ }\textbf {\bibinfo {volume} {4}},\ \bibinfo {pages} {40}
  (\bibinfo {year} {2021}{\natexlab{b}})}\BibitemShut {NoStop}%
\bibitem [{\citenamefont {Haga}\ \emph {et~al.}(2022)\citenamefont {Haga},
  \citenamefont {Opletal}, \citenamefont {Tokiwa}, \citenamefont {Yamamoto},
  \citenamefont {Tokunaga}, \citenamefont {Kambe},\ and\ \citenamefont
  {Sakai}}]{Haga2022}%
  \BibitemOpen
  \bibfield  {author} {\bibinfo {author} {\bibfnamefont {Y.}~\bibnamefont
  {Haga}}, \bibinfo {author} {\bibfnamefont {P.}~\bibnamefont {Opletal}},
  \bibinfo {author} {\bibfnamefont {Y.}~\bibnamefont {Tokiwa}}, \bibinfo
  {author} {\bibfnamefont {E.}~\bibnamefont {Yamamoto}}, \bibinfo {author}
  {\bibfnamefont {Y.}~\bibnamefont {Tokunaga}}, \bibinfo {author}
  {\bibfnamefont {S.}~\bibnamefont {Kambe}}, \ and\ \bibinfo {author}
  {\bibfnamefont {H.}~\bibnamefont {Sakai}},\ }\href {\doibase
  10.1088/1361-648X/ac5201} {\bibfield  {journal} {\bibinfo  {journal} {J.
  Phys.: Condens. Matter}\ }\textbf {\bibinfo {volume} {34}},\ \bibinfo {pages}
  {175601} (\bibinfo {year} {2022})}\BibitemShut {NoStop}%
\bibitem [{\citenamefont {Sakai}\ \emph {et~al.}(2022)\citenamefont {Sakai},
  \citenamefont {Opletal}, \citenamefont {Tokiwa}, \citenamefont {Yamamoto},
  \citenamefont {Tokunaga}, \citenamefont {Kambe},\ and\ \citenamefont
  {Haga}}]{Sakai2022}%
  \BibitemOpen
  \bibfield  {author} {\bibinfo {author} {\bibfnamefont {H.}~\bibnamefont
  {Sakai}}, \bibinfo {author} {\bibfnamefont {P.}~\bibnamefont {Opletal}},
  \bibinfo {author} {\bibfnamefont {Y.}~\bibnamefont {Tokiwa}}, \bibinfo
  {author} {\bibfnamefont {E.}~\bibnamefont {Yamamoto}}, \bibinfo {author}
  {\bibfnamefont {Y.}~\bibnamefont {Tokunaga}}, \bibinfo {author}
  {\bibfnamefont {S.}~\bibnamefont {Kambe}}, \ and\ \bibinfo {author}
  {\bibfnamefont {Y.}~\bibnamefont {Haga}},\ }\href {\doibase
  10.1103/PhysRevMaterials.6.073401} {\bibfield  {journal} {\bibinfo  {journal}
  {Phys. Rev. Materials}\ }\textbf {\bibinfo {volume} {6}},\ \bibinfo {pages}
  {073401} (\bibinfo {year} {2022})}\BibitemShut {NoStop}%
\bibitem [{\citenamefont {Tokiwa}\ \emph {et~al.}()\citenamefont {Tokiwa},
  \citenamefont {Opletal}, \citenamefont {Sakai}, \citenamefont {Kubo},
  \citenamefont {Yamamoto}, \citenamefont {Kambe}, \citenamefont {Kimata},
  \citenamefont {Awaji}, \citenamefont {Sasaki}, \citenamefont {Aoki},
  \citenamefont {Tokunaga},\ and\ \citenamefont {Haga}}]{Tokiwa_arxiv}%
  \BibitemOpen
  \bibfield  {author} {\bibinfo {author} {\bibfnamefont {Y.}~\bibnamefont
  {Tokiwa}}, \bibinfo {author} {\bibfnamefont {P.}~\bibnamefont {Opletal}},
  \bibinfo {author} {\bibfnamefont {H.}~\bibnamefont {Sakai}}, \bibinfo
  {author} {\bibfnamefont {K.}~\bibnamefont {Kubo}}, \bibinfo {author}
  {\bibfnamefont {E.}~\bibnamefont {Yamamoto}}, \bibinfo {author}
  {\bibfnamefont {S.}~\bibnamefont {Kambe}}, \bibinfo {author} {\bibfnamefont
  {M.}~\bibnamefont {Kimata}}, \bibinfo {author} {\bibfnamefont
  {S.}~\bibnamefont {Awaji}}, \bibinfo {author} {\bibfnamefont
  {T.}~\bibnamefont {Sasaki}}, \bibinfo {author} {\bibfnamefont
  {D.}~\bibnamefont {Aoki}}, \bibinfo {author} {\bibfnamefont {Y.}~\bibnamefont
  {Tokunaga}}, \ and\ \bibinfo {author} {\bibfnamefont {Y.}~\bibnamefont
  {Haga}},\ }\href@noop {} {\bibinfo  {journal} {arXiv:2210.11769}\
  }\BibitemShut {NoStop}%
\bibitem [{\citenamefont {Aoki}\ \emph
  {et~al.}(2022{\natexlab{b}})\citenamefont {Aoki}, \citenamefont {Sakai},
  \citenamefont {Opletal}, \citenamefont {Tokiwa}, \citenamefont {Ishizuka},
  \citenamefont {Yanase}, \citenamefont {Harima}, \citenamefont {Nakamura},
  \citenamefont {Li}, \citenamefont {Homma}, \citenamefont {Shimizu},
  \citenamefont {Knebel}, \citenamefont {Flouquet},\ and\ \citenamefont
  {Haga}}]{Aoki2022}%
  \BibitemOpen
\bibfield  {journal} {  }\bibfield  {author} {\bibinfo {author} {\bibfnamefont
  {D.}~\bibnamefont {Aoki}}, \bibinfo {author} {\bibfnamefont {H.}~\bibnamefont
  {Sakai}}, \bibinfo {author} {\bibfnamefont {P.}~\bibnamefont {Opletal}},
  \bibinfo {author} {\bibfnamefont {Y.}~\bibnamefont {Tokiwa}}, \bibinfo
  {author} {\bibfnamefont {J.}~\bibnamefont {Ishizuka}}, \bibinfo {author}
  {\bibfnamefont {Y.}~\bibnamefont {Yanase}}, \bibinfo {author} {\bibfnamefont
  {H.}~\bibnamefont {Harima}}, \bibinfo {author} {\bibfnamefont
  {A.}~\bibnamefont {Nakamura}}, \bibinfo {author} {\bibfnamefont
  {D.}~\bibnamefont {Li}}, \bibinfo {author} {\bibfnamefont {Y.}~\bibnamefont
  {Homma}}, \bibinfo {author} {\bibfnamefont {Y.}~\bibnamefont {Shimizu}},
  \bibinfo {author} {\bibfnamefont {G.}~\bibnamefont {Knebel}}, \bibinfo
  {author} {\bibfnamefont {J.}~\bibnamefont {Flouquet}}, \ and\ \bibinfo
  {author} {\bibfnamefont {Y.}~\bibnamefont {Haga}},\ }\href {\doibase
  10.7566/JPSJ.91.083704} {\bibfield  {journal} {\bibinfo  {journal} {J. Phys.
  Soc. Jpn.}\ }\textbf {\bibinfo {volume} {91}},\ \bibinfo {pages} {083704}
  (\bibinfo {year} {2022}{\natexlab{b}})}\BibitemShut {NoStop}%
\bibitem [{\citenamefont {Tanaka}\ \emph {et~al.}(2022)\citenamefont {Tanaka},
  \citenamefont {Mizukami}, \citenamefont {Harasawa}, \citenamefont
  {Hashimoto}, \citenamefont {Hwang}, \citenamefont {Kurita}, \citenamefont
  {Tanaka}, \citenamefont {Fujimoto}, \citenamefont {Matsuda}, \citenamefont
  {Moon},\ and\ \citenamefont {Shibauchi}}]{Tanaka2022}%
  \BibitemOpen
  \bibfield  {author} {\bibinfo {author} {\bibfnamefont {O.}~\bibnamefont
  {Tanaka}}, \bibinfo {author} {\bibfnamefont {Y.}~\bibnamefont {Mizukami}},
  \bibinfo {author} {\bibfnamefont {R.}~\bibnamefont {Harasawa}}, \bibinfo
  {author} {\bibfnamefont {K.}~\bibnamefont {Hashimoto}}, \bibinfo {author}
  {\bibfnamefont {K.}~\bibnamefont {Hwang}}, \bibinfo {author} {\bibfnamefont
  {N.}~\bibnamefont {Kurita}}, \bibinfo {author} {\bibfnamefont
  {H.}~\bibnamefont {Tanaka}}, \bibinfo {author} {\bibfnamefont
  {S.}~\bibnamefont {Fujimoto}}, \bibinfo {author} {\bibfnamefont
  {Y.}~\bibnamefont {Matsuda}}, \bibinfo {author} {\bibfnamefont {E.-G.}\
  \bibnamefont {Moon}}, \ and\ \bibinfo {author} {\bibfnamefont
  {T.}~\bibnamefont {Shibauchi}},\ }\href {\doibase 10.1038/s41567-021-01488-6}
  {\bibfield  {journal} {\bibinfo  {journal} {Nat. Phys.}\ }\textbf {\bibinfo
  {volume} {18}},\ \bibinfo {pages} {429} (\bibinfo {year} {2022})}\BibitemShut
  {NoStop}%
\bibitem [{\citenamefont {Shibauchi}\ \emph {et~al.}(2007)\citenamefont
  {Shibauchi}, \citenamefont {Konczykowski}, \citenamefont {van~der Beek},
  \citenamefont {Okazaki}, \citenamefont {Matsuda}, \citenamefont {Yamaura},
  \citenamefont {Nagao},\ and\ \citenamefont {Hiroi}}]{Shibauchi2007}%
  \BibitemOpen
  \bibfield  {author} {\bibinfo {author} {\bibfnamefont {T.}~\bibnamefont
  {Shibauchi}}, \bibinfo {author} {\bibfnamefont {M.}~\bibnamefont
  {Konczykowski}}, \bibinfo {author} {\bibfnamefont {C.~J.}\ \bibnamefont
  {van~der Beek}}, \bibinfo {author} {\bibfnamefont {R.}~\bibnamefont
  {Okazaki}}, \bibinfo {author} {\bibfnamefont {Y.}~\bibnamefont {Matsuda}},
  \bibinfo {author} {\bibfnamefont {J.}~\bibnamefont {Yamaura}}, \bibinfo
  {author} {\bibfnamefont {Y.}~\bibnamefont {Nagao}}, \ and\ \bibinfo {author}
  {\bibfnamefont {Z.}~\bibnamefont {Hiroi}},\ }\href {\doibase
  10.1103/PhysRevLett.99.257001} {\bibfield  {journal} {\bibinfo  {journal}
  {Phys. Rev. Lett.}\ }\textbf {\bibinfo {volume} {99}},\ \bibinfo {pages}
  {257001} (\bibinfo {year} {2007})}\BibitemShut {NoStop}%
\bibitem [{\citenamefont {Okazaki}\ \emph {et~al.}(2009)\citenamefont
  {Okazaki}, \citenamefont {Konczykowski}, \citenamefont {van~der Beek},
  \citenamefont {Kato}, \citenamefont {Hashimoto}, \citenamefont {Shimozawa},
  \citenamefont {Shishido}, \citenamefont {Yamashita}, \citenamefont
  {Ishikado}, \citenamefont {Kito}, \citenamefont {Iyo}, \citenamefont
  {Eisaki}, \citenamefont {Shamoto}, \citenamefont {Shibauchi},\ and\
  \citenamefont {Matsuda}}]{Okazaki2009}%
  \BibitemOpen
  \bibfield  {author} {\bibinfo {author} {\bibfnamefont {R.}~\bibnamefont
  {Okazaki}}, \bibinfo {author} {\bibfnamefont {M.}~\bibnamefont
  {Konczykowski}}, \bibinfo {author} {\bibfnamefont {C.~J.}\ \bibnamefont
  {van~der Beek}}, \bibinfo {author} {\bibfnamefont {T.}~\bibnamefont {Kato}},
  \bibinfo {author} {\bibfnamefont {K.}~\bibnamefont {Hashimoto}}, \bibinfo
  {author} {\bibfnamefont {M.}~\bibnamefont {Shimozawa}}, \bibinfo {author}
  {\bibfnamefont {H.}~\bibnamefont {Shishido}}, \bibinfo {author}
  {\bibfnamefont {M.}~\bibnamefont {Yamashita}}, \bibinfo {author}
  {\bibfnamefont {M.}~\bibnamefont {Ishikado}}, \bibinfo {author}
  {\bibfnamefont {H.}~\bibnamefont {Kito}}, \bibinfo {author} {\bibfnamefont
  {A.}~\bibnamefont {Iyo}}, \bibinfo {author} {\bibfnamefont {H.}~\bibnamefont
  {Eisaki}}, \bibinfo {author} {\bibfnamefont {S.}~\bibnamefont {Shamoto}},
  \bibinfo {author} {\bibfnamefont {T.}~\bibnamefont {Shibauchi}}, \ and\
  \bibinfo {author} {\bibfnamefont {Y.}~\bibnamefont {Matsuda}},\ }\href
  {\doibase 10.1103/PhysRevB.79.064520} {\bibfield  {journal} {\bibinfo
  {journal} {Phys. Rev. B}\ }\textbf {\bibinfo {volume} {79}},\ \bibinfo
  {pages} {064520} (\bibinfo {year} {2009})}\BibitemShut {NoStop}%
\bibitem [{\citenamefont {Okazaki}\ \emph {et~al.}(2010)\citenamefont
  {Okazaki}, \citenamefont {Shimozawa}, \citenamefont {Shishido}, \citenamefont
  {Konczykowski}, \citenamefont {Haga}, \citenamefont {D.~Matsuda},
  \citenamefont {Yamamoto}, \citenamefont {Onuki}, \citenamefont {Yanase},
  \citenamefont {Shibauchi},\ and\ \citenamefont {Matsuda}}]{Okazaki2010}%
  \BibitemOpen
  \bibfield  {author} {\bibinfo {author} {\bibfnamefont {R.}~\bibnamefont
  {Okazaki}}, \bibinfo {author} {\bibfnamefont {M.}~\bibnamefont {Shimozawa}},
  \bibinfo {author} {\bibfnamefont {H.}~\bibnamefont {Shishido}}, \bibinfo
  {author} {\bibfnamefont {M.}~\bibnamefont {Konczykowski}}, \bibinfo {author}
  {\bibfnamefont {Y.}~\bibnamefont {Haga}}, \bibinfo {author} {\bibfnamefont
  {T.}~\bibnamefont {D.~Matsuda}}, \bibinfo {author} {\bibfnamefont
  {E.}~\bibnamefont {Yamamoto}}, \bibinfo {author} {\bibfnamefont
  {Y.}~\bibnamefont {Onuki}}, \bibinfo {author} {\bibfnamefont
  {Y.}~\bibnamefont {Yanase}}, \bibinfo {author} {\bibfnamefont
  {T.}~\bibnamefont {Shibauchi}}, \ and\ \bibinfo {author} {\bibfnamefont
  {Y.}~\bibnamefont {Matsuda}},\ }\href {\doibase 10.1143/JPSJ.79.084705}
  {\bibfield  {journal} {\bibinfo  {journal} {J. Phys. Soc. Jpn.}\ }\textbf
  {\bibinfo {volume} {79}},\ \bibinfo {pages} {084705} (\bibinfo {year}
  {2010})}\BibitemShut {NoStop}%
\bibitem [{\citenamefont {Putzke}\ \emph {et~al.}(2014)\citenamefont {Putzke},
  \citenamefont {Walmsley}, \citenamefont {Fletcher}, \citenamefont {Malone},
  \citenamefont {Vignolles}, \citenamefont {Proust}, \citenamefont {Badoux},
  \citenamefont {See}, \citenamefont {Beere}, \citenamefont {Ritchie},
  \citenamefont {Kasahara}, \citenamefont {Mizukami}, \citenamefont
  {Shibauchi}, \citenamefont {Matsuda},\ and\ \citenamefont
  {Carrington}}]{Putzke2014}%
  \BibitemOpen
  \bibfield  {author} {\bibinfo {author} {\bibfnamefont {C.}~\bibnamefont
  {Putzke}}, \bibinfo {author} {\bibfnamefont {P.}~\bibnamefont {Walmsley}},
  \bibinfo {author} {\bibfnamefont {J.~D.}\ \bibnamefont {Fletcher}}, \bibinfo
  {author} {\bibfnamefont {L.}~\bibnamefont {Malone}}, \bibinfo {author}
  {\bibfnamefont {D.}~\bibnamefont {Vignolles}}, \bibinfo {author}
  {\bibfnamefont {C.}~\bibnamefont {Proust}}, \bibinfo {author} {\bibfnamefont
  {S.}~\bibnamefont {Badoux}}, \bibinfo {author} {\bibfnamefont
  {P.}~\bibnamefont {See}}, \bibinfo {author} {\bibfnamefont {H.~E.}\
  \bibnamefont {Beere}}, \bibinfo {author} {\bibfnamefont {D.~A.}\ \bibnamefont
  {Ritchie}}, \bibinfo {author} {\bibfnamefont {S.}~\bibnamefont {Kasahara}},
  \bibinfo {author} {\bibfnamefont {Y.}~\bibnamefont {Mizukami}}, \bibinfo
  {author} {\bibfnamefont {T.}~\bibnamefont {Shibauchi}}, \bibinfo {author}
  {\bibfnamefont {Y.}~\bibnamefont {Matsuda}}, \ and\ \bibinfo {author}
  {\bibfnamefont {A.}~\bibnamefont {Carrington}},\ }\href {\doibase
  10.1038/ncomms6679} {\bibfield  {journal} {\bibinfo  {journal} {Nat.
  Commun.}\ }\textbf {\bibinfo {volume} {5}},\ \bibinfo {pages} {5679}
  (\bibinfo {year} {2014})}\BibitemShut {NoStop}%
\bibitem [{\citenamefont {Paulsen}\ \emph {et~al.}(2021)\citenamefont
  {Paulsen}, \citenamefont {Knebel}, \citenamefont {Lapertot}, \citenamefont
  {Braithwaite}, \citenamefont {Pourret}, \citenamefont {Aoki}, \citenamefont
  {Hardy}, \citenamefont {Flouquet},\ and\ \citenamefont
  {Brison}}]{Paulsen2021}%
  \BibitemOpen
  \bibfield  {author} {\bibinfo {author} {\bibfnamefont {C.}~\bibnamefont
  {Paulsen}}, \bibinfo {author} {\bibfnamefont {G.}~\bibnamefont {Knebel}},
  \bibinfo {author} {\bibfnamefont {G.}~\bibnamefont {Lapertot}}, \bibinfo
  {author} {\bibfnamefont {D.}~\bibnamefont {Braithwaite}}, \bibinfo {author}
  {\bibfnamefont {A.}~\bibnamefont {Pourret}}, \bibinfo {author} {\bibfnamefont
  {D.}~\bibnamefont {Aoki}}, \bibinfo {author} {\bibfnamefont {F.}~\bibnamefont
  {Hardy}}, \bibinfo {author} {\bibfnamefont {J.}~\bibnamefont {Flouquet}}, \
  and\ \bibinfo {author} {\bibfnamefont {J.-P.}\ \bibnamefont {Brison}},\
  }\href {\doibase 10.1103/PhysRevB.103.L180501} {\bibfield  {journal}
  {\bibinfo  {journal} {Phys. Rev. B}\ }\textbf {\bibinfo {volume} {103}},\
  \bibinfo {pages} {L180501} (\bibinfo {year} {2021})}\BibitemShut {NoStop}%
\bibitem [{\citenamefont {Brandt}(1999)}]{Brandt1999}%
  \BibitemOpen
  \bibfield  {author} {\bibinfo {author} {\bibfnamefont {E.~H.}\ \bibnamefont
  {Brandt}},\ }\href {\doibase 10.1103/PhysRevB.60.11939} {\bibfield  {journal}
  {\bibinfo  {journal} {Phys. Rev. B}\ }\textbf {\bibinfo {volume} {60}},\
  \bibinfo {pages} {11939} (\bibinfo {year} {1999})}\BibitemShut {NoStop}%
\bibitem [{\citenamefont {Kittaka}\ \emph {et~al.}(2020)\citenamefont
  {Kittaka}, \citenamefont {Shimizu}, \citenamefont {Sakakibara}, \citenamefont
  {Nakamura}, \citenamefont {Li}, \citenamefont {Homma}, \citenamefont {Honda},
  \citenamefont {Aoki},\ and\ \citenamefont {Machida}}]{Kittaka2020}%
  \BibitemOpen
  \bibfield  {author} {\bibinfo {author} {\bibfnamefont {S.}~\bibnamefont
  {Kittaka}}, \bibinfo {author} {\bibfnamefont {Y.}~\bibnamefont {Shimizu}},
  \bibinfo {author} {\bibfnamefont {T.}~\bibnamefont {Sakakibara}}, \bibinfo
  {author} {\bibfnamefont {A.}~\bibnamefont {Nakamura}}, \bibinfo {author}
  {\bibfnamefont {D.}~\bibnamefont {Li}}, \bibinfo {author} {\bibfnamefont
  {Y.}~\bibnamefont {Homma}}, \bibinfo {author} {\bibfnamefont
  {F.}~\bibnamefont {Honda}}, \bibinfo {author} {\bibfnamefont
  {D.}~\bibnamefont {Aoki}}, \ and\ \bibinfo {author} {\bibfnamefont
  {K.}~\bibnamefont {Machida}},\ }\href {\doibase
  10.1103/PhysRevResearch.2.032014} {\bibfield  {journal} {\bibinfo  {journal}
  {Phys. Rev. Research}\ }\textbf {\bibinfo {volume} {2}},\ \bibinfo {pages}
  {032014(R)} (\bibinfo {year} {2020})}\BibitemShut {NoStop}%
\bibitem [{\citenamefont {Song}\ \emph {et~al.}(2011)\citenamefont {Song},
  \citenamefont {Ghim}, \citenamefont {Yoon}, \citenamefont {Lee},
  \citenamefont {Jung}, \citenamefont {Ji}, \citenamefont {Shim}, \citenamefont
  {Bang},\ and\ \citenamefont {Kwon}}]{Song2011}%
  \BibitemOpen
  \bibfield  {author} {\bibinfo {author} {\bibfnamefont {Y.~J.}\ \bibnamefont
  {Song}}, \bibinfo {author} {\bibfnamefont {J.~S.}\ \bibnamefont {Ghim}},
  \bibinfo {author} {\bibfnamefont {J.~H.}\ \bibnamefont {Yoon}}, \bibinfo
  {author} {\bibfnamefont {K.~J.}\ \bibnamefont {Lee}}, \bibinfo {author}
  {\bibfnamefont {M.~H.}\ \bibnamefont {Jung}}, \bibinfo {author}
  {\bibfnamefont {H.-S.}\ \bibnamefont {Ji}}, \bibinfo {author} {\bibfnamefont
  {J.~H.}\ \bibnamefont {Shim}}, \bibinfo {author} {\bibfnamefont
  {Y.}~\bibnamefont {Bang}}, \ and\ \bibinfo {author} {\bibfnamefont {Y.~S.}\
  \bibnamefont {Kwon}},\ }\href {\doibase 10.1209/0295-5075/94/57008}
  {\bibfield  {journal} {\bibinfo  {journal} {EPL}\ }\textbf {\bibinfo {volume}
  {94}},\ \bibinfo {pages} {57008} (\bibinfo {year} {2011})}\BibitemShut
  {NoStop}%
\bibitem [{\citenamefont {Adamski}\ \emph {et~al.}(2017)\citenamefont
  {Adamski}, \citenamefont {Krellner},\ and\ \citenamefont
  {Abdel-Hafiez}}]{Adamski2017}%
  \BibitemOpen
  \bibfield  {author} {\bibinfo {author} {\bibfnamefont {A.}~\bibnamefont
  {Adamski}}, \bibinfo {author} {\bibfnamefont {C.}~\bibnamefont {Krellner}}, \
  and\ \bibinfo {author} {\bibfnamefont {M.}~\bibnamefont {Abdel-Hafiez}},\
  }\href {\doibase 10.1103/PhysRevB.96.100503} {\bibfield  {journal} {\bibinfo
  {journal} {Phys. Rev. B}\ }\textbf {\bibinfo {volume} {96}},\ \bibinfo
  {pages} {100503(R)} (\bibinfo {year} {2017})}\BibitemShut {NoStop}%
\bibitem [{\citenamefont {Vincent}\ \emph {et~al.}(1991)\citenamefont
  {Vincent}, \citenamefont {Hammann}, \citenamefont {Taillefer}, \citenamefont
  {Behnia}, \citenamefont {Keller},\ and\ \citenamefont
  {Flouquet}}]{Vincent1991}%
  \BibitemOpen
  \bibfield  {author} {\bibinfo {author} {\bibfnamefont {E.}~\bibnamefont
  {Vincent}}, \bibinfo {author} {\bibfnamefont {J.}~\bibnamefont {Hammann}},
  \bibinfo {author} {\bibfnamefont {L.}~\bibnamefont {Taillefer}}, \bibinfo
  {author} {\bibfnamefont {K.}~\bibnamefont {Behnia}}, \bibinfo {author}
  {\bibfnamefont {N.}~\bibnamefont {Keller}}, \ and\ \bibinfo {author}
  {\bibfnamefont {J.}~\bibnamefont {Flouquet}},\ }\href {\doibase
  10.1088/0953-8984/3/20/013} {\bibfield  {journal} {\bibinfo  {journal} {J.
  Phys.: Condens. Matter}\ }\textbf {\bibinfo {volume} {3}},\ \bibinfo {pages}
  {3517} (\bibinfo {year} {1991})}\BibitemShut {NoStop}%
\bibitem [{\citenamefont {Amann}\ \emph {et~al.}(1998)\citenamefont {Amann},
  \citenamefont {Mota}, \citenamefont {Maple},\ and\ \citenamefont
  {v.~L\"ohneysen}}]{Amann1998}%
  \BibitemOpen
  \bibfield  {author} {\bibinfo {author} {\bibfnamefont {A.}~\bibnamefont
  {Amann}}, \bibinfo {author} {\bibfnamefont {A.~C.}\ \bibnamefont {Mota}},
  \bibinfo {author} {\bibfnamefont {M.~B.}\ \bibnamefont {Maple}}, \ and\
  \bibinfo {author} {\bibfnamefont {H.}~\bibnamefont {vonL\"ohneysen}},\ }\href
  {\doibase 10.1103/PhysRevB.57.3640} {\bibfield  {journal} {\bibinfo
  {journal} {Phys. Rev. B}\ }\textbf {\bibinfo {volume} {57}},\ \bibinfo
  {pages} {3640} (\bibinfo {year} {1998})}\BibitemShut {NoStop}%
\bibitem [{\citenamefont {Cichorek}\ \emph {et~al.}(2005)\citenamefont
  {Cichorek}, \citenamefont {Mota}, \citenamefont {Steglich}, \citenamefont
  {Frederick}, \citenamefont {Yuhasz},\ and\ \citenamefont
  {Maple}}]{Cichorek2005}%
  \BibitemOpen
  \bibfield  {author} {\bibinfo {author} {\bibfnamefont {T.}~\bibnamefont
  {Cichorek}}, \bibinfo {author} {\bibfnamefont {A.~C.}\ \bibnamefont {Mota}},
  \bibinfo {author} {\bibfnamefont {F.}~\bibnamefont {Steglich}}, \bibinfo
  {author} {\bibfnamefont {N.~A.}\ \bibnamefont {Frederick}}, \bibinfo {author}
  {\bibfnamefont {W.~M.}\ \bibnamefont {Yuhasz}}, \ and\ \bibinfo {author}
  {\bibfnamefont {M.~B.}\ \bibnamefont {Maple}},\ }\href {\doibase
  10.1103/PhysRevLett.94.107002} {\bibfield  {journal} {\bibinfo  {journal}
  {Phys. Rev. Lett.}\ }\textbf {\bibinfo {volume} {94}},\ \bibinfo {pages}
  {107002} (\bibinfo {year} {2005})}\BibitemShut {NoStop}%
\bibitem [{\citenamefont {Chia}\ \emph {et~al.}(2003)\citenamefont {Chia},
  \citenamefont {Salamon}, \citenamefont {Sugawara},\ and\ \citenamefont
  {Sato}}]{Chia2003}%
  \BibitemOpen
  \bibfield  {author} {\bibinfo {author} {\bibfnamefont {E.~E.~M.}\
  \bibnamefont {Chia}}, \bibinfo {author} {\bibfnamefont {M.~B.}\ \bibnamefont
  {Salamon}}, \bibinfo {author} {\bibfnamefont {H.}~\bibnamefont {Sugawara}}, \
  and\ \bibinfo {author} {\bibfnamefont {H.}~\bibnamefont {Sato}},\ }\href
  {\doibase 10.1103/PhysRevLett.91.247003} {\bibfield  {journal} {\bibinfo
  {journal} {Phys. Rev. Lett.}\ }\textbf {\bibinfo {volume} {91}},\ \bibinfo
  {pages} {247003} (\bibinfo {year} {2003})}\BibitemShut {NoStop}%
\bibitem [{\citenamefont {Iguchi}\ \emph {et~al.}()\citenamefont {Iguchi},
  \citenamefont {Man}, \citenamefont {Thomas}, \citenamefont {Ronning},
  \citenamefont {Rosa},\ and\ \citenamefont {Moler}}]{Iguchi_arxiv}%
  \BibitemOpen
  \bibfield  {author} {\bibinfo {author} {\bibfnamefont {Y.}~\bibnamefont
  {Iguchi}}, \bibinfo {author} {\bibfnamefont {H.}~\bibnamefont {Man}},
  \bibinfo {author} {\bibfnamefont {S.~M.}\ \bibnamefont {Thomas}}, \bibinfo
  {author} {\bibfnamefont {F.}~\bibnamefont {Ronning}}, \bibinfo {author}
  {\bibfnamefont {P.~F.~S.}\ \bibnamefont {Rosa}}, \ and\ \bibinfo {author}
  {\bibfnamefont {K.~A.}\ \bibnamefont {Moler}},\ }\href@noop {} {\bibinfo
  {journal} {arXiv:2210.09562}\ }\BibitemShut {NoStop}%
\bibitem [{\citenamefont {Nakamine}\ \emph {et~al.}(2019)\citenamefont
  {Nakamine}, \citenamefont {Kitagawa}, \citenamefont {Ishida}, \citenamefont
  {Tokunaga}, \citenamefont {Sakai}, \citenamefont {Kambe}, \citenamefont
  {Nakamura}, \citenamefont {Shimizu}, \citenamefont {Homma}, \citenamefont
  {Li}, \citenamefont {Honda},\ and\ \citenamefont {Aoki}}]{Nakamine2019}%
  \BibitemOpen
\bibfield  {journal} {  }\bibfield  {author} {\bibinfo {author} {\bibfnamefont
  {G.}~\bibnamefont {Nakamine}}, \bibinfo {author} {\bibfnamefont
  {S.}~\bibnamefont {Kitagawa}}, \bibinfo {author} {\bibfnamefont
  {K.}~\bibnamefont {Ishida}}, \bibinfo {author} {\bibfnamefont
  {Y.}~\bibnamefont {Tokunaga}}, \bibinfo {author} {\bibfnamefont
  {H.}~\bibnamefont {Sakai}}, \bibinfo {author} {\bibfnamefont
  {S.}~\bibnamefont {Kambe}}, \bibinfo {author} {\bibfnamefont
  {A.}~\bibnamefont {Nakamura}}, \bibinfo {author} {\bibfnamefont
  {Y.}~\bibnamefont {Shimizu}}, \bibinfo {author} {\bibfnamefont
  {Y.}~\bibnamefont {Homma}}, \bibinfo {author} {\bibfnamefont
  {D.}~\bibnamefont {Li}}, \bibinfo {author} {\bibfnamefont {F.}~\bibnamefont
  {Honda}}, \ and\ \bibinfo {author} {\bibfnamefont {D.}~\bibnamefont {Aoki}},\
  }\href {\doibase 10.7566/JPSJ.88.113703} {\bibfield  {journal} {\bibinfo
  {journal} {J. Phys. Soc. Jpn.}\ }\textbf {\bibinfo {volume} {88}},\ \bibinfo
  {pages} {113703} (\bibinfo {year} {2019})}\BibitemShut {NoStop}%
\bibitem [{\citenamefont {Sundar}\ \emph {et~al.}()\citenamefont {Sundar},
  \citenamefont {Azari}, \citenamefont {Goeks}, \citenamefont {Gheidi},
  \citenamefont {Abedi}, \citenamefont {Yakovlev}, \citenamefont {Dunsiger},
  \citenamefont {Wilkinson}, \citenamefont {Blundell}, \citenamefont {Metz},
  \citenamefont {Hayes}, \citenamefont {Saha}, \citenamefont {Lee},
  \citenamefont {Woods}, \citenamefont {Movshovich}, \citenamefont {Thomas},
  \citenamefont {Rosa}, \citenamefont {Butch}, \citenamefont {Paglione},\ and\
  \citenamefont {Sonier}}]{Sundar_arxiv}%
  \BibitemOpen
  \bibfield  {author} {\bibinfo {author} {\bibfnamefont {S.}~\bibnamefont
  {Sundar}}, \bibinfo {author} {\bibfnamefont {N.}~\bibnamefont {Azari}},
  \bibinfo {author} {\bibfnamefont {M.}~\bibnamefont {Goeks}}, \bibinfo
  {author} {\bibfnamefont {S.}~\bibnamefont {Gheidi}}, \bibinfo {author}
  {\bibfnamefont {M.}~\bibnamefont {Abedi}}, \bibinfo {author} {\bibfnamefont
  {M.}~\bibnamefont {Yakovlev}}, \bibinfo {author} {\bibfnamefont {S.~R.}\
  \bibnamefont {Dunsiger}}, \bibinfo {author} {\bibfnamefont {J.~M.}\
  \bibnamefont {Wilkinson}}, \bibinfo {author} {\bibfnamefont {S.~J.}\
  \bibnamefont {Blundell}}, \bibinfo {author} {\bibfnamefont {T.~E.}\
  \bibnamefont {Metz}}, \bibinfo {author} {\bibfnamefont {I.~M.}\ \bibnamefont
  {Hayes}}, \bibinfo {author} {\bibfnamefont {S.~R.}\ \bibnamefont {Saha}},
  \bibinfo {author} {\bibfnamefont {S.}~\bibnamefont {Lee}}, \bibinfo {author}
  {\bibfnamefont {A.~J.}\ \bibnamefont {Woods}}, \bibinfo {author}
  {\bibfnamefont {R.}~\bibnamefont {Movshovich}}, \bibinfo {author}
  {\bibfnamefont {S.~M.}\ \bibnamefont {Thomas}}, \bibinfo {author}
  {\bibfnamefont {P.~F.~S.}\ \bibnamefont {Rosa}}, \bibinfo {author}
  {\bibfnamefont {N.~P.}\ \bibnamefont {Butch}}, \bibinfo {author}
  {\bibfnamefont {J.}~\bibnamefont {Paglione}}, \ and\ \bibinfo {author}
  {\bibfnamefont {J.~E.}\ \bibnamefont {Sonier}},\ }\href@noop {} {\bibinfo
  {journal} {arXiv:2207.13725}\ }\BibitemShut {NoStop}%
\bibitem [{\citenamefont {Hashimoto}\ \emph {et~al.}(2012)\citenamefont
  {Hashimoto}, \citenamefont {Cho}, \citenamefont {Shibauchi}, \citenamefont
  {Kasahara}, \citenamefont {Mizukami}, \citenamefont {Katsumata},
  \citenamefont {Tsuruhara}, \citenamefont {Terashima}, \citenamefont {Ikeda},
  \citenamefont {Tanatar}, \citenamefont {Kitano}, \citenamefont {Salovich},
  \citenamefont {Giannetta}, \citenamefont {Walmsley}, \citenamefont
  {Carrington}, \citenamefont {Prozorov},\ and\ \citenamefont
  {Matsuda}}]{Hashimoto2012}%
  \BibitemOpen
\bibfield  {journal} {  }\bibfield  {author} {\bibinfo {author} {\bibfnamefont
  {K.}~\bibnamefont {Hashimoto}}, \bibinfo {author} {\bibfnamefont
  {K.}~\bibnamefont {Cho}}, \bibinfo {author} {\bibfnamefont {T.}~\bibnamefont
  {Shibauchi}}, \bibinfo {author} {\bibfnamefont {S.}~\bibnamefont {Kasahara}},
  \bibinfo {author} {\bibfnamefont {Y.}~\bibnamefont {Mizukami}}, \bibinfo
  {author} {\bibfnamefont {R.}~\bibnamefont {Katsumata}}, \bibinfo {author}
  {\bibfnamefont {Y.}~\bibnamefont {Tsuruhara}}, \bibinfo {author}
  {\bibfnamefont {T.}~\bibnamefont {Terashima}}, \bibinfo {author}
  {\bibfnamefont {H.}~\bibnamefont {Ikeda}}, \bibinfo {author} {\bibfnamefont
  {M.~A.}\ \bibnamefont {Tanatar}}, \bibinfo {author} {\bibfnamefont
  {H.}~\bibnamefont {Kitano}}, \bibinfo {author} {\bibfnamefont
  {N.}~\bibnamefont {Salovich}}, \bibinfo {author} {\bibfnamefont {R.~W.}\
  \bibnamefont {Giannetta}}, \bibinfo {author} {\bibfnamefont {P.}~\bibnamefont
  {Walmsley}}, \bibinfo {author} {\bibfnamefont {A.}~\bibnamefont
  {Carrington}}, \bibinfo {author} {\bibfnamefont {R.}~\bibnamefont
  {Prozorov}}, \ and\ \bibinfo {author} {\bibfnamefont {Y.}~\bibnamefont
  {Matsuda}},\ }\href {\doibase 10.1126/science.1219821} {\bibfield  {journal}
  {\bibinfo  {journal} {Science}\ }\textbf {\bibinfo {volume} {336}},\ \bibinfo
  {pages} {1554} (\bibinfo {year} {2012})}\BibitemShut {NoStop}%
\end{thebibliography}
\end{document}